\documentclass[twocolumn]{aastex631} 
\usepackage{newtxtext,newtxmath}
\usepackage[T1]{fontenc}   

\DeclareRobustCommand{\VAN}[3]{#2}
\let\VANthebibliography\thebibliography
\def\thebibliography{\DeclareRobustCommand{\VAN}[3]{##3}\VANthebibliography}
\usepackage{graphicx}
\usepackage{multirow}
\usepackage[utf8]{inputenc}
\usepackage{graphicx,color}
\usepackage{mathtools}
\usepackage{amsmath}
\usepackage{booktabs}
\usepackage{upgreek}
\usepackage{bm}
\usepackage{pstricks} 
\usepackage[normalem]{ulem}
\usepackage{cancel}
    \usepackage[british]{babel}

\def\aap{A\&A}
\def\apjl{ApJ}

\def\apjs{ApJS}

\newcommand{\f}{_\textrm{0}}

\newcommand{\del}{\partial}

\newcommand{\bfDel}{\bm{\nabla}}

\newcommand{\bfB}{\bm{B}}

\newcommand{\bfu}{\bm{u}}

\newcommand{\bfb}{\bm{b}}
\newcommand{\bmb}{\bm{b}}

\newcommand{\mean}[1]{\overline{#1}}
\newcommand{\meanv}[1]{\overline{\bm{#1}}}

\newcommand{\D}{_\mathrm{D}}						
\newcommand{\eq}{_\mathrm{eq}}						
\newcommand{\z}{_\mathrm{0}}					   	
\newcommand{\kin}{_\mathrm{k}}			   		
\newcommand{\magn}{_\mathrm{m}}			   		
\newcommand{\turb}{_\mathrm{tur}}			   		
\newcommand{\crit}{_\mathrm{c}}			   		
\newcommand{\const}{\text{const}}			   	
\newcommand{\dd}{\mathrm{d}}			   		
\newcommand{\pol}{_\mathrm{pol}}			   		

\newcommand{\cro}{\times}
\newcommand{\Rm}{\mathrm{Rm}}

\newcommand{\rtilde}{\widetilde{r}}

\newcommand{\Btilde}{\widetilde{B}}

\newcommand{\Sigmatilde}{\widetilde{\Sigma}}

\newcommand{\I}{_\mathrm{I}}

\newcommand{\Ma}{\mathcal{M}}
\newcommand{\Pm}{\mathrm{Pm}}

\newcommand{\Msun}{\,\mathrm{M_\odot}}

\newcommand{\Msunyrkpckpc}{\,\mathrm{M_\odot\,yr^{-1}\kpc^{-2}}}
\newcommand{\Msunkpckpc}{\,\mathrm{M_\odot\,\kpc^{-2}}}
\newcommand{\Msunpcpc}{\,\mathrm{M_\odot\,\pc^{-2}}}

\newcommand{\tot}{_\mathrm{tot}}

\newcommand{\A}{_\mathrm{A}}
\newcommand{\B}{_\mathrm{B}}

\newcommand{\HI}{H$\,$\textsc{i} }

\newcommand{\mH}{m_\mathrm{H}}

\newcommand{\w}{_\mathrm{wd}}

\newcommand{\thm}{_\mathrm{thm}}

\newcommand{\Mach}{\mathcal{M}}

\newcommand{\sfr}{_\mathrm{SFR}}

\newcommand{\SN}{_\mathrm{SN}}

\newcommand{\SB}{_\mathrm{SB}}
\newcommand{\uSN}{^\mathrm{SN}}
\newcommand{\uSB}{^\mathrm{SB}}
\newcommand{\sound}{_\mathrm{s}}

\newcommand{\blowout}{_\mathrm{b}}

\newcommand{\ren}{^\mathrm{r}}
\newcommand{\eddy}{_\mathrm{e}}
\newcommand{\renov}{_\mathrm{r}}
\newcommand{\taur}{\tau\renov}
\newcommand{\taue}{\tau\eddy}
\newcommand{\E}{E}

\newcommand{\inj}{^\mathrm{i}}

\newcommand{\Bbar}{\mean{B}}
\newcommand{\St}{\Sigma\tot}
\newcommand{\Sf}{\Sigma\sfr}
\newcommand{\Ss}{\Sigma_\star}
\newcommand{\Sg}{\Sigma}

\newcommand{\Om}{\Omega}
\newcommand{\kB}{k_\mathrm{B}}
\newcommand{\iso}{_\mathrm{iso}}
\newcommand{\ani}{_\mathrm{ani}}

\newcommand{\Vc}{V_\mathrm{c}}

\definecolor{perhaps_teal}{rgb}{0, 0.5, 0.5}

\definecolor{mediumorchid}{rgb}{0.73, 0.33, 0.83}

\newcommand{\cm}{\,{\rm cm}}

\newcommand{\cmcube}{\,{\rm cm^{-3}}}

\newcommand{\erg}{\,{\rm erg}}
\newcommand{\g}{\,{\rm g}}
\newcommand{\gcmcmcm}{\,{\rm g\,cm^{-3}}}

\newcommand{\kms}{\,{\rm km\,s^{-1}}}
\newcommand{\kmskpc}{\,{\rm km\,s^{-1}\,kpc^{-1}}}

\newcommand{\K}{\,{\rm K}}
\newcommand{\kpc}{\,{\rm kpc}}
\newcommand{\pc}{\,{\rm pc}}
\newcommand{\Mpc}{\,{\rm Mpc}}
\newcommand{\Myr}{\,{\rm Myr}}

\newcommand{\mkG}{\,\upmu{\rm G}}

\newcommand{\p}{\,{\rm pc}}

\newcommand{\s}{\,{\rm s}}
\newcommand{\yr}{\,{\rm yr}}

\definecolor{mediumorchid}{rgb}{0.73, 0.33, 0.83}

\begin{document}

\title{Galactic magnetic fields I. Theoretical model and scaling relations}

\author{Luke Chamandy}
\affiliation{National Institute of Science Education and Research, An OCC of Homi Bhabha National Institute, Bhubaneswar 752050, Odisha, India}
\affiliation{Department of Physics and Astronomy, University of Rochester, Rochester NY 14627, USA}

\author{Rion Glenn Nazareth}
\affiliation{National Institute of Science Education and Research, An OCC of Homi Bhabha National Institute, Bhubaneswar 752050, Odisha, India}

\author{Gayathri Santhosh}
\affiliation{National Institute of Science Education and Research, An OCC of Homi Bhabha National Institute, Bhubaneswar 752050, Odisha, India}

\correspondingauthor{Luke Chamandy}
\email{lchamandy@niser.ac.in}
\begin{abstract}
Galactic dynamo models have generally relied on input parameters that are very challenging to constrain.  We address this problem by developing a model that uses observable quantities as input: the galaxy rotation curve, the surface densities of the gas, stars and star formation rate, and the gas temperature.  The model can be used to estimate parameters of the random and mean components of the magnetic field, as well as the gas scale height, root-mean-square velocity and the correlation length and time of the interstellar turbulence, in terms of the observables.  We use our model to derive theoretical scaling relations for the quantities of interest, finding reasonable agreement with empirical scaling relations inferred from observation.  We assess the dependence of the results on different assumptions about turbulence driving, finding that agreement with observations is improved by explicitly modeling the expansion and energetics of supernova remnants.  The model is flexible enough to include alternative prescriptions for the physical processes involved, and we provide links to two open-source \textsc{python} programs that implement it.
\end{abstract}

%
%
\section{Introduction}
\label{sec:intro}
Spiral galaxies generally host magnetic fields with strengths of order $10\mkG$.
The general properties of such fields can be explained using turbulent dynamo models, 
but detailed comparison between theory and observation is still rudimentary
\citep[for reviews, see][]{Ruzmaikin+88,Beck+96,Shukurov04,Shukurov05,Vaneck+15,Beck+19,Shukurov+Subramanian21}. 
The situation can be improved using various 
complementary approaches. 
One of them is to evaluate observable quantities using simple, 
preferably analytic models of galactic magnetic fields, 
since they are readily available, transparent, easy to apply and can be used to explore the parameter space. 
However, such models tend to depend on poorly known galactic parameters,
such as the turbulent correlation time $\tau$ and length $l$. 
This problem can now be partially circumvented by using a model for the turbulence parameters \citep{Chamandy+Shukurov20} 
whose inputs can be expressed in terms of observable quantities.

The point of the present work is to write down a closed set of coupled algebraic equations 
that can be solved to 
obtain magnetic field properties from a handful of galactic observables,
to use these equations to derive scaling relations in certain asymptotic limits,
and to compare these relations with those suggested by observations.%
\footnote{Scaling relations are relations of the form $X\propto x^ay^bz^c\cdots$, 
where $X$ is a quantity of interest,
$x$, $y$ and $z$ are observable quantities, and $a$, $b$ and $c$ are constants.}
We make no attempt to include cosmological evolution,
and thus our model is restricted to low-redshift galaxies 
(redshift $\ll1$).

Magnetic fields can play a role in various aspects of galactic astrophysics
given that their energy densities are comparable to those of turbulence, 
cosmic rays and thermal motions.
Scaling relations can be particularly useful in modeling such effects.
Take, for instance, 
the empirical scaling relation between the total magnetic field strength $B$
and star formation rate (SFR) surface density $\Sf$ (Section~\ref{sec:j_disk}).
The exponent in the relation is a key ingredient in the model of \citet{Wang+Lilly22}, 
which aims to explain the origin of the exponential profiles of the disks of star-forming galaxies.
In that model, radial gas inflow in the disk is driven by the Maxwell stress,
which depends on the magnetic field strength, 
which is in turn linked to the gas surface density $\Sg$ 
through the star formation law governing the relation between $\Sg$ and $\Sf$. 

Furthermore, magnetic fields can be used as tracers of physical processes like star formation.
Consider that the scaling relation between $B$ and $\Sf$
is also useful for modeling the infrared-radio correlation of star-forming galaxies
\citep[][and references therein]{Schober+23}.
Here, radio emission is dominated by synchrotron and far infrared 
by thermal emission from dust.
A physical link is provided by star formation, 
which leads to the cosmic ray electrons involved in synchrotron 
and the heating of the dust.

While empirical scaling relations can be used directly in physical models,
a truly self-consistent model requires such relations to be derived using physical arguments.
This is one of the key motivations for the present work.

The paper is organized as follows. Section~\ref{sec:model} presents an overview of the model. 
Then we present each part of the model separately: the magnetic field in Section~\ref{sec:magnetic_field}, 
the turbulence parameters in Section~\ref{sec:turbulence}, 
and the remaining relevant quantities in Section~\ref{sec:observables}. 
In Section~\ref{sec:scaling} we show how 
solutions can be approximated as scaling relations,
and compare these to observations and numerical simulations. 
Finally, we summarize our results and provide conclusions in Section~\ref{sec:conclusions}.

\section{Overview of the model}
\label{sec:model}

The model combines existing analytic solutions for the mean magnetic field \citep{Chamandy+14b}
and parameters of the supernova-driven interstellar turbulence \citep{Chamandy+Shukurov20}.
Those solutions have been tested against 
asymptotic and numerical solutions in the case of mean-field dynamo models \citep{Chamandy+14b,Chamandy16,Chamandy+16},
and against direct numerical simulations 
of the local interstellar medium in the case of turbulence models \citep{Hollins+17}.
Furthermore, local galactic dynamo simulations that solve the full equations of MHD are
reasonably consistent with dynamo theory \citep{Gressel+08a,Gressel+08b,Gent+23,Gent+24}.
In addition to our mean-field dynamo and interstellar turbulence models, 
we use suitably motivated heuristic expressions 
for quantities like the random magnetic field strength and gas scale height. 
For simplicity, the gaseous disk is approximated as being comprised of a single-phase fluid in a statistical steady state.
We use cylindrical polar coordinates ($r,\phi,z$) 
with the $z$-axis aligned with the galaxy's rotation axis
and $z=0$ at the galactic midplane.
All variables depend on the galactocentric radius $r$ and represent vertical  
and azimuthal averages over the gaseous galactic disk
(although we discuss the azimuthal variations in connection with the effects of the spiral arms).

The model is summarized in Fig.~\ref{fig:flowchart_conceptual}.
The observables on which the turbulence and magnetic field properties depend are 
the angular speed of the gas rotation about the galactic centre $\Omega(r)$, 
the dimensionless rotational velocity shear rate 
\begin{equation}\label{q}
  q\equiv-\frac{\dd\ln\Omega}{\dd\ln r}\,, 
\end{equation}
the stellar surface mass density $\Sigma_\star(r)$, 
the gas surface mass density $\Sigma(r)$, 
the surface density of the SFR $\Sigma\sfr(r)$
and the gas temperature $T(r)$. 
Note that $q>0$ since $\dd\Omega/\dd r<0$ in most parts of galaxies, 
and $q=1$ for a flat (constant circular velocity) rotation curve.
These input quantities -- with the exception of $\Omega(r)$ and $q(r)$ -- are 
first used to calculate the gas scale height $h(r)$,
the gas density $\rho(r)$, the supernova (SN) rate density $\nu(r)$, 
the root-mean-square (rms) turbulent velocity $u(r)$,
the turbulent correlation length $l(r)$ and turbulent correlation time $\tau(r)$.
These quantities are, in turn, used as input for the magnetic field model, 
along with $\Omega(r)$ and $q(r)$.
The magnetic field parameters include  
the rms strength of the random (turbulent) magnetic field $b(r)$, 
its degree of anisotropy (defined below),
the strength $\mean{B}(r)$ of the mean magnetic field,
and its pitch angle 
\begin{equation}
  \tan p_B(r)= \frac{\mean{B}_r(r)}{\mean{B}_\phi(r)},
\end{equation}
where $-\pi/2<p_B\le\pi/2$.

The key quantities in the model are defined and summarized in Table~\ref{tab:params}. 
Source code has been provided for quickly obtaining scaling relations
in various asymptotic limits,
including but not limited to the ones presented in Section~\ref{sec:scaling}.%
\footnote{See \url{https://github.com/Rnazx/Scaling-Relations}.}
In Paper~II, we will apply our model to specific galaxies.
For this purpose, we will solve the general equations using semi-analytic methods.%
\footnote{The source code is available at \url{https://github.com/Rnazx/goblin}.} 

\begin{table*}
  \begin{center}
  \caption{Key quantities in the model. 
           Adjustable parameters are either fixed at the value in the table, or, 
           as in the case of $\psi$, 
           allowed to vary within a small range between galaxies but not within them.
          \label{tab:params}
          }
  \renewcommand{\arraystretch}{0.95} 
  \begin{tabular}{@{}ccccc@{}}
\hline
Type	&Symbol		&Quantity						&Typical units or value		&Reference  \\
\hline
Observable       &$d$		&Galaxy distance						&$\!\Mpc$		&Measured			\\
"       &$\rtilde$		&Galactocentric radius (angular)			&$'$		&"			\\
"       &$\Vc(r)$		&Circular speed						&$\!\kms$		&"			\\
"       &$\Sigma(r)$	&Gas surface mass density				&$\!\Msunkpckpc$	&"			\\
"       &$\Ss(r)$ or $\St(r)$	&Stellar or total mass surface density		&$\!\Msunkpckpc$	&"			\\
"       &$\Sigma\sfr(r)$	&Star formation rate surface density			&$\!\Msunyrkpckpc$	&"			\\
"       &$T(r)$		&Gas temperature					&$\!\K$			&"			\\
\hline
Derived Observable       &$r$		&Galactocentric radius						&$\!\kpc$		&$\rtilde d$			\\
"\phantom{ble para}"       &$\Omega(r)$	&Gas galactocentric angular speed			&$\!\kmskpc$		&$\Omega=\Vc/r$			\\
"\phantom{ble para}"       &$q(r)$		&Radial shear parameter				&--			&$q\equiv-\dd\ln\Omega/\dd\ln r$			\\
\hline
Modelled       &$\varepsilon(r)$		&Degree of anisotropy of random magnetic field 		&--		&Eq.~\eqref{varepsilon} 	\\
"       &$b\ani(r)$		&Anisotropic random magnetic field strength 		&$\!\mkG$		&Eq.~\eqref{b_ani_noUz} or Eq.~\eqref{b_ani_gen} 	\\
"       &$b\iso(r)$		&Isotropic random magnetic field strength 		&$\!\mkG$		&Eq.~\eqref{b_iso}		\\
"       &$B\eq(r)$		&Equipartition field strength				&$\!\gcmcmcm$		&Eq.~\eqref{Beq}		\\
"       &$\Bbar(r)$		&Mean magnetic field strength  				&$\!\mkG$		&Eq.~\eqref{Bbar_noUz} or Eq.~\eqref{Bbar_gen} 	\\
"       &$D\crit$	&Critical dynamo number 				&--			&Eq.~\eqref{Dc_noUz} or Eq.~\eqref{Dc_gen} 			\\
"       &$D(r)$ 	&Dynamo number 
&--			&Eq.~\eqref{Dk}			\\
"       &$\eta(r)$		&Turbulent diffusivity of $\bm{\Bbar}$			&$\!\cm^2\s^{-1}$	&Eq.~\eqref{eta}		\\
"       &$\alpha\magn(r)$	&$\alpha$ effect term from nonlinear backreaction	&$\!\kms$		&Eq.~\eqref{alpham}		\\
"       &$\alpha\kin(r)$	&$\alpha$ effect in kinematic regime 			&$\!\kms$		&Eq.~\eqref{alphak_approx} 	\\
"       &$p_B(r)$		&Mean magnetic field pitch angle  			&$^\circ$		&Eq.~\eqref{p_B_noUz} or Eq.~\eqref{p_B_gen} 	\\
"       &$l(r)$		&Turbulent correlation length scale			&$\!\kpc$		&Eq.~\eqref{l_noSBs} or Eq.~\eqref{l_gen} \\
"       &$l\SN(r)$		&Driving scale of turbulence by isolated SNe				&$\!\kpc$		&Eq.~\eqref{l_SN}		\\
"       &$n(r)$		&Ambient gas number density				&$\!\cmcube$		&Eq.~\eqref{n}			\\
"       &$u(r)$		&Root-mean-square turbulent velocity			&$\!\kms$		&Eq.~\eqref{u_noSBs} or Eq.~\eqref{u_gen} \\
"       &$\tau(r)$		&Turbulent correlation time				&$\!\Myr$		&Eq.~\eqref{tau}		\\
"       &$\tau\eddy(r)$	&Turnover time of energy-carrying eddies		&$\!\Myr$		&Eq.~\eqref{tau_eddy}		\\
"       &$\tau\renov(r)$	&Renovation time of turbulence			&$\!\Myr$		&Eq.~\eqref{tau_renov_noSBs}  or Eq.~\eqref{tau_renov_gen}	\\
"       &$\nu(r)$        	&SN rate per unit volume                   		&$\!\kpc^{-3}\Myr^{-1}$ &Eq.~\eqref{nu}			\\
"       &$\rho(r)$		&Gas mass density					&$\!\gcmcmcm$		&Eq.~\eqref{rho}	    	\\
"       &$c\sound(r)$	&Sound speed						&$\!\kms$		&Eq.~\eqref{cs}		\\
"       &$\Mach(r)$		&Turbulent sonic Mach number $u/c\sound$		&--		&Eq.~\eqref{Mach}	\\
"       &$h(r)$		&Gas scale height					&$\!\kpc$		&Eq.~\eqref{h}		\\
\hline
Constant       &$\xi\f$		&Ratio $b\iso^2/B\eq^2$ for $\Mach\le A_1$			&$0.4$			&Eq.~\eqref{b_iso}			\\
"       &$A_1$		&Value of $\Mach$ above which $b\iso$ reduces	   &$\sqrt{2}$			&Eq.~\eqref{b_iso}			\\
"       &$R_\kappa$	&Ratio $\kappa/\eta$ of turbulent diffusivity of $\alpha\magn$ to that of $\meanv{B}$	&$0.3$	&Eq.~\eqref{Bbar_noUz} or \eqref{Bbar_gen} \\
"       &$C'_\alpha $	&Coefficient that sets $\alpha\kin$ ceiling		&$1$			&Eq.~\eqref{alphak_approx}		\\
"       &$C_l$		&Constant from turbulence theory					&$3/4$			&Eq.~\eqref{l_noSBs} or \eqref{l_gen}	\\
"       &$\Gamma$	&Turbulence spectral index				&$5/3$			&Eq.~\eqref{l_noSBs} or \eqref{l_gen}		\\
"       &$E\SN$ 	    	&Initial SN energy		     			&$10^{51}\erg$ 		&Eq.~\eqref{l_SN}			\\
"       &$\delta$	&Fraction of stars that evolve to SNe			&$8\times10^{-3}$	&Eq.~\eqref{nu}	\\
"       &$m_\star$	&Average stellar mass					&$0.85\Msun$		&Eq.~\eqref{nu}	\\
"       &$A_2$		&Sets relative strength of thermal pressure term	   &$\sqrt{2}$			&Eq.~\eqref{w}			\\
"       &$N$		&Exponent in Kennicutt-Schmidt law			&$1.4$       		&Eq.~\eqref{KS}	\\
\hline
Adjustable parameter       &$\beta$     &Multiplies magnetic field strengths    &-- &Eq.~\eqref{Beq}		\\
"\phantom{ble para}"       &$K$		&Normalization factor for $\Bbar$	&-- &Eq.~\eqref{Bbar_noUz} or \eqref{Bbar_gen}\\
"\phantom{ble para}"       &$C_\alpha$	&Numerical coefficient of $\alpha\kin$ effect	&-- &Eq.~\eqref{alphak_approx}	\\
"\phantom{ble para}"       &$\psi$	&Multiplies the isolated SN driving scale	&--   &Eq.~\eqref{l_SN}	\\
"\phantom{ble para}"       &$\zeta$     &Factor in expression for gas scale height     &-- &Eq.~\eqref{h}		\\
    \hline                  	 
  \end{tabular}
  \end{center}
  \renewcommand{\arraystretch}{1.0} 
\end{table*}

\section{Magnetic field model}
\label{sec:magnetic_field}
Dynamo amplification is fast enough that the magnetic fields of nearby galaxies (definitely their random components and, likely, their large-scale components, at least in the central parts of galaxies)
are expected to be in a saturated state \citep{Beck+94,Rodrigues+19}. 
In this statistically steady state, 
they are predicted to have magnetic energy density 
similar to the turbulent kinetic energy density
and this is borne out in observations \citep{Beck+19}. 
Thus, in this work we consider only the saturated state of the magnetic field.
We separate the field into two main components, 
the mean field $\meanv{B}$ and the random field $\bmb$, 
so that we may write the total field as 
\begin{equation}
  \label{B}
  \bfB=\meanv{B}+\bmb,
\end{equation}
where overbar represents a suitable averaging 
(spatial in the case of observed variables and ensemble 
averages in theoretical results).
This separation is both physically and mathematically appropriate 
because the two terms in equation~\eqref{B} are believed to be governed by distinct physical processes 
and are sensitive to different parameters
\citep[e.g.,][]{Brandenburg+Subramanian05a,Shukurov+Subramanian21}, 
and also because they have distinct observational signatures.

\subsection{Random field}
\label{sec:random_field}
Magnetic fields in galaxies are inferred from observations 
to usually be dominated by a random small-scale component \citep{Beck+19}.
Such random fields can be produced by fluctuation (small-scale) dynamo action,
as well as the turbulent tangling of the mean magnetic field.
It is also possible that part of the random component detected in observations
is actually unresolved mean field.
Below we neglect the influence of the mean field on the random field for simplicity,
leaving such effects for future study.

\begin{figure*}
  \includegraphics[width=1.0\textwidth, clip=true, trim= 50 350 100 100]{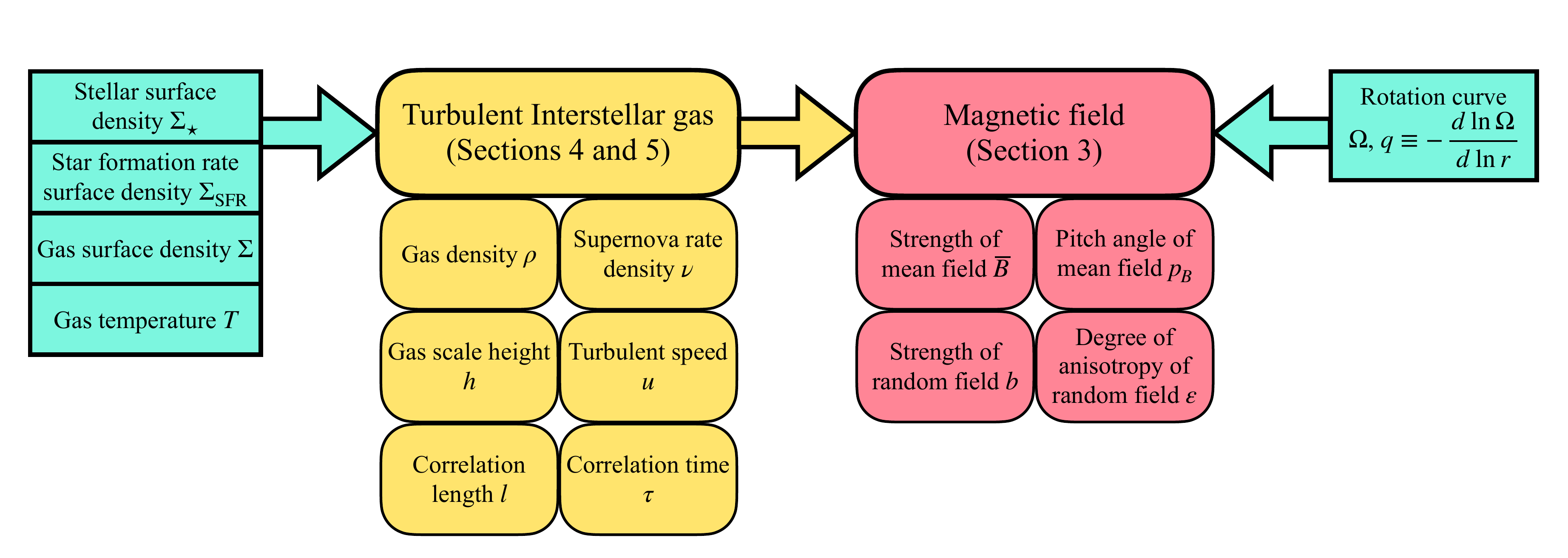}
  \vspace{-0.5cm}
  \caption{The structure of the model showing, with reference to various parts of the text, the hierarchy of input and derived parameters.}
  \label{fig:flowchart_conceptual}
\end{figure*}

Random galactic fields are inferred from observations to be anisotropic \citep[e.g.][]{Fletcher+11}.
Below, we assume that this anisotropy is solely produced 
by the large-scale galactic shear arising from differential rotation.
Other sources of anisotropy, 
such as shear and compression in spiral density waves, could also be important.
The galactic differential rotation stretches the radial component of the random field,
leading to a linear increase with time of the azimuthal component.
This can lead to a cumulative effect for a duration of about the correlation time of turbulence $\tau$.
Let the standard deviation of any component of the random field be given by $\sigma_i=\sqrt{\mean{b_i^2}}$, with $i=(r,\phi,z)$.
For a random magnetic field in a statistically steady (saturated) state,
we can estimate \citep[][\S 4.4.1]{Shukurov+Subramanian21}
\begin{equation}
  \label{sigma_phi}
  \sigma_\phi= \sigma_r(1+q\Omega\tau).
\end{equation}
Galactic outflows (winds or fountain flow) may also contain large-scale shear, 
which would enhance $\sigma_z$ relative to $\sigma_r$. 
We neglect galactic outflows for simplicity, 
but general expressions that include the mean vertical outflow speed $U\f$ are given in Appendix~\ref{sec:outflow}.
With $U\f=0$, we obtain 
\begin{equation}\label{sigma_z_noUz}
  \sigma_z=\sigma_r.
\end{equation}
The strength of the random field is thus given by 
\begin{equation}\label{b}
  b\equiv \sqrt{\mean{\bmb^2}}= \sqrt{\sigma_r^2+\sigma_\phi^2+\sigma_z^2}=\sigma_r\sqrt{2+(1+q\Omega\tau)^2}
\end{equation}
and that of the isotropic background by 
\begin{equation}\label{b_iso_1}
  b\iso= \sqrt{3}\sigma_r.
\end{equation}
The degree of anisotropy can be expressed as
\begin{equation}
\label{varepsilon}
  \varepsilon=\frac{b\ani}{b},
\end{equation}
where 
\begin{equation}
\label{b_ani_noUz}
  b\ani\equiv \sqrt{b^2-b\iso^2}= \frac{b\iso}{\sqrt{3}}\left[2q\Omega\tau\left(1+\frac{q\Omega\tau}{2}\right)\right]^{1/2}.
\end{equation}

Simulations of fluctuation dynamos in periodic boxes with isotropic forcing 
can be used to estimate the strength of the random field in the saturated state.
Since large-scale shear is typically not included in such simulations, 
they are best used to estimate $b\iso$, rather than $b$.
In particular, they can help to constrain the ratio $\xi$ of energy density of the saturated magnetic field
$b\iso^2/8\uppi$ to that of the turbulent motions $\tfrac{1}{2}\rho u^2$, 
for a range of magnetic Reynolds and magnetic Prandtl numbers;
in galaxies, one expects $\Rm\gg1$ and $\Pm\gg1$.
These simulations suggest that $\xi=b\iso^2/4\uppi\rho u^2$ ranges from $\approx0.4$ for solenoidally forced subsonic turbulence,
to values that are considerably smaller for compressively forced or supersonic turbulence. 
Based loosely on results from \citet{Federrath+11} (see also \citealt{Seta+Federrath21}) we choose
\begin{equation}
  \label{b_iso}
  b\iso= \frac{\xi\f^{1/2} B\eq}{\max(1,\Mach/A_1)},
\end{equation}
where $A_1$ is constant of order unity,
\begin{equation}\label{Mach}
\Mach\equiv \frac{u}{c\sound}
\end{equation}
is the turbulent sonic Mach number, 
$c\sound$ is the speed of sound and $\xi\f=0.4$.
Here, 
\begin{equation}
  \label{Beq}
  B\eq= \beta(4\uppi\rho)^{1/2} u
\end{equation}
is the field strength corresponding to energy equipartition with turbulence.
A parameter, $\beta$, has been inserted to account for 
imprecision in both theory and observational inference.

\subsection{Mean field}\label{sec:mean_field}
Spiral galaxies also contain magnetic field components 
that are coherent on scales up to the system size.
This large-scale component (sometimes called the mean or regular field)
can generally be explained by appealing to mean-field dynamo action \citep{Beck+19}.
Below we apply $\alpha$-$\Omega$ dynamo theory including a nonlinear 
backreaction that quenches the dynamo and leads to saturation
\citep{Pouquet+76,Krause+Radler80,Kleeorin+Ruzmaikin82,Ruzmaikin+88,
Blackman+Field02,Radler+03,Brandenburg+Subramanian05a,Shukurov+06,
Subramanian+Brandenburg06,Sur+07b,Chamandy+14b,Gopalakrishnan+Subramanian23}.

We make use of an analytic steady-state 
solution of the mean-field dynamo equations
that assumes a thin disc
and employs the no-$z$ approximation 
to replace $z$-derivatives by divisions by $h(r)$, 
with coefficients chosen to produce good agreement with solutions 
that retain the $z$-dependence
\citep{Subramanian+Mestel93,Moss95,Phillips01,Chamandy+14b}.
General expressions that allow for a finite mean vertical velocity 
are derived in \citet{Chamandy+14b} and can be found in Section~\ref{sec:outflow}.
In the absence of mean vertical outflow or inflow, 
the strength of the mean magnetic field in the saturated state is given by
\begin{equation}
  \label{Bbar_noUz}
  \Bbar\equiv |\bm{\Bbar}|= K\frac{\pi l}{h}\left[\left(\frac{D}{D\crit} -1\right)R_\kappa\right]^{1/2}B\eq,
\end{equation}
where $K$ is a factor of order $0.1$--$1$ 
that accounts for theoretical uncertainty \citep{Chamandy+Singh18},
$D$ is the dynamo number, 
subscripts `k' and `c' refer to the kinematic ($\Bbar\ll B\eq$) and critical 
(no growth or decay) values,
and $D>D\crit$ (supercritical dynamo); 
if $D\le D\crit$ then we instead adopt $\Bbar=0$.
The critical dynamo number is given by
\begin{equation}
  \label{Dc_noUz}
  D\crit= -\left(\frac{\uppi}{2}\right)^5,
\end{equation}
and the kinematic dynamo number by
\begin{equation}
  \label{Dk}
  D= R_\alpha R_\Omega,
\end{equation}
with Reynolds-type dimensionless numbers
\begin{equation}
  \label{Reynolds}
  R_\alpha\equiv \frac{\alpha\kin h}{\eta}, \qquad R_\Omega\equiv -\frac{q\Omega h^2}{\eta}, 
               \qquad R_\kappa\equiv \frac{\kappa}{\eta}.
\end{equation}
where
\begin{equation}\label{alphak}
  \alpha\kin = -\frac{1}{3}\tau\,\overline{\bfu\cdot\bfDel\cro \bfu},
\end{equation}
the turbulent diffusivity of $\meanv{B}$ is given by 
\begin{equation}
  \label{eta}
  \eta= \frac{1}{3}\tau u^2,
\end{equation}
and $\kappa$ is the turbulent diffusivity of the quantity 
\begin{equation}\label{alpham}
  \alpha\magn= \frac{1}{3}\tau\frac{\mean{\bfb\cdot\bfDel\cro\bfb}}{4\uppi\rho},
\end{equation}
which becomes important in the nonlinear regime owing to the backreaction 
of the field onto the flow.
The total contribution to the $\alpha$ effect is obtained by summing $\alpha\kin$ and $\alpha\magn$.
We approximate $\alpha\kin$ as \citep{Ruzmaikin+88,Chamandy+16}
\begin{equation}
  \label{alphak_approx}
  \alpha\kin= 
      \begin{dcases}
        \frac{C_\alpha\tau^2u^2\Omega}{h},
        \quad  &\mbox{if  } \min\left(1,\frac{C'_\alpha h}{C_\alpha\tau u}\right)\ge \Omega\tau;\\
        \frac{C_\alpha\tau u^2}{h},
        &\mbox{if  } \min\left(\Omega\tau,\frac{C'_\alpha h}{C_\alpha\tau u}\right)\ge1;\\
        C'_\alpha u,
        &\mbox{if  } \min\left(\Omega\tau,1\right)\ge\frac{C'_\alpha h}{C_\alpha\tau u},
      \end{dcases}
\end{equation}
where $C_\alpha$ and $C'_\alpha $ are constants of order unity 
that account for a lack of precision in the modeling;
note that $C'_\alpha u$ acts as a ceiling on $\alpha\kin$.
The pitch angle of $\bm{\Bbar}$ is given by 
\begin{equation}
  \label{p_B_noUz}
  \tan p_B= \frac{\Bbar_r}{\Bbar_\phi}
  = \frac{\uppi^2}{4R_\Omega} 
  = -\frac{\uppi^2\,\tau\,u^2}{12\,q\,\Omega\, h^2},
\end{equation}
defined such that $-\uppi/2<p_B\le\uppi/2$ with $p_B<0$ for trailing spirals (with respect to the galactic rotation).

\section{Interstellar turbulence model}
\label{sec:turbulence}
Magnetic field solutions depend on the rms turbulent velocity $u$, 
the turbulent correlation time $\tau$ and length $l$, 
which are either very challenging ($u$ and $l$) 
or impossible ($\tau$) to measure directly from observations.
The quantity $u$ can be inferred from the line-of-sight velocity dispersion, 
which is commonly observed in galaxies,
but the data is contaminated by contributions from thermal motions, cloud motions, and outflows,
which are difficult to separate out \citep[e.g.][]{Mogotsi+16}.%
\footnote{See \citealt{Burkhart21} for a review of methods to study interstellar turbulence using observations.}
Therefore, we seek solutions for these turbulence quantities in terms of accessible observable quantities.

Supernova (SN) feedback is generally believed to be the dominant driver of turbulence 
in nearby galaxies \citep[e.g.][]{Klessen+Glover16,Krumholz+18,Bacchini+20}.
For SN-driven turbulence, $u$, $\tau$ and $l$ have been estimated using an
analytic model that considers turbulence to be simultaneously driven by isolated SNe 
and superbubbles (SBs) formed by multiple SNe in OB associations \citep{Chamandy+Shukurov20}.
In the model, a fraction $f\SB$ of SNe are assumed to contribute to SBs. 
Here we neglect SBs by setting $f\SB=0$, 
but we include a summary of the general model ($f\SB\ne0$) in Appendix~\ref{sec:SBs}.

\subsection{Correlation length}\label{sec:l}
For $f\SB=0$, equation~\eqref{l_gen} for the turbulent correlation length becomes
\begin{equation}
  \label{l_noSBs}
  l= \left(\frac{\Gamma-1}{\Gamma}\right)C_l l\SN=\frac{3}{10}l\SN,
\end{equation}
where we have assumed $\Gamma=5/3$ and $C_l=3/4$ (see Section~\ref{sec:l_gen} and Table~\ref{tab:params}),
and where $l\SN$ is the driving scale of turbulence driven by isolated SNe.
The quantity $l\SN$ is equal to the radius of the SN remnant (SNR) when its age is $t\sound\uSN$, 
defined as the time at which the SNR expansion speed becomes equal to the ambient sound speed, 
$\dot{R}\SN= c\sound$.
At this time the SNR is assumed to fragment and merge with the interstellar medium (ISM), 
transferring its energy to the latter. 
This gives
\begin{equation}
  \label{l_SN}
    l\SN= R\SN(t\sound\uSN)= 0.14\kpc\; \psi E_{51}^{16/51}n_{0.1}^{-19/51}c_{10}^{-1/3},
\end{equation}
where $E_{51}=E\SN/10^{51}\erg$ is the SN energy (excluding that in neutrinos), 
$n_{0.1}=n/0.1\cmcube$ is the gas number density, 
and $c_{10}=c\sound/10\kms$.
A dimensionless parameter of order unity, $\psi$, 
is introduced in the present work 
to account for the uncertainty in the model.
To convert from mass density to number density, 
we have used the expression
\begin{equation}
  \label{n}
  n= \frac{\rho}{\mu\mH},
\end{equation}
with mean molecular mass $\mu=14/11$.

\subsection{Root-mean-square turbulent velocity}\label{sec:u}
The rms turbulent velocity, $u$, 
is estimated by equating the energy injection rate per unit volume 
from SNe with the dissipation rate per unit volume $\sim \rho u^3/2l$ 
owing to the turbulent energy cascade.
With $f\SB=0$, equation~\eqref{u_gen} becomes
\begin{equation}
  \label{u_noSBs}
  u= \left( \frac{4\uppi}{3}l\SN^3l c\sound^2\nu 
     \right)^{1/3},
\end{equation}
with $l$ given by equation~\eqref{l_noSBs} and $l\SN$ by equation~\eqref{l_SN}.

\subsection{Correlation time}\label{sec:tau}
We estimate the correlation time $\tau$ as the 
minimum of the turnover time $\tau\eddy$ of energy-carrying eddies,
and the time $\tau\renov$ for the flow to renovate due to the passage of an SN blast wave,
\begin{equation}
  \label{tau}
  \tau=\min(\tau\renov,\tau\eddy).
\end{equation}
The eddy turnover time (comparable to the lifetime of the largest eddies) 
is estimated as 
\begin{equation}
  \label{tau_eddy}
  \tau\eddy=\frac{l}{u},
\end{equation}
where $l$ is given by equation~\eqref{l_noSBs} and $u$ is given by equation~\eqref{u_noSBs} for the case $f\SB=0$.
For this case, the renovation time is given by
\begin{equation}
  \label{tau_renov_noSBs}
    \tau\renov= \tau\SN\ren= \frac{3}{4\uppi l\SN^3\nu}= 6.8\Myr\; \frac{1}{4}\nu_{50}^{-1}E_{51}^{-16/17}n_{0.1}^{19/17}c_{10}.
\end{equation}
Equations~\eqref{l_noSBs} (or \ref{l_gen}), \eqref{u_noSBs} (or \ref{u_gen}) and \eqref{tau},
can be used wherever the quantities $l$, $u$ and $\tau$ appear 
in the equations of Section~\ref{sec:magnetic_field}.

\section{Formulating the equations in terms of the observables}
\label{sec:observables}
We must still obtain expressions for the SN rate per unit volume $\nu(r)$, 
gas density $\rho(r)$, 
disk scale height $h(r)$ and 
sound speed $c\sound(r)$ in terms of the observable quantities. 

Following \citet{Vaneck+15}, we write
\begin{equation}
  \label{nu}
  \nu= \frac{\delta\Sigma\sfr}{2hm_\star}
\end{equation}
where $\Sigma\sfr/2h$ is the mean SFR volume density averaged across the gas disk,
$\delta\approx8\times10^{-3}$ is the fraction of stars that evolve to SNe
for the initial mass function (IMF) of \citet{Kroupa08}, 
and $m_\star=0.85\Msun$ is the average stellar mass for this IMF
(other choices of IMF would lead to small changes in $\delta$ and $m_\star$).

For an exponential or uniform density disk of scale height or half-thickness $h$, 
we have
\begin{equation}
  \label{rho}
  \rho= \frac{\Sg}{2h}.
\end{equation}

Approximating the ISM as a uniform ideal gas, we write the sound speed as
\begin{equation}
  \label{cs}
  c\sound= \left(\frac{\gamma \kB T}{\mu\mH}\right)^{1/2}.
\end{equation}
A reasonable choice for the adiabatic index is $\gamma=1.5$ \citep{Vandenbroucke+13}.

The scale height $h$ can be estimated from vertical hydrostatic balance \citep[e.g.][]{Rodrigues+19}.
We use
\begin{equation}
  \label{h}
  h\approx \frac{w^2}{3\uppi G(\Sg+\Ss/\zeta)} 
  \approx \zeta\frac{w^2}{3\pi G\St},
\end{equation}
where 
\begin{equation}
  \label{w}
  w\equiv(u^2+A_2^2c\sound^2)^{1/2}.
\end{equation} 
Here $A_2^2=2$ for $\gamma=3/2$,
$\Ss$ is the surface density of stars, $\sigma_\star$ is the 1D velocity dispersion of stars,
$\zeta$ (formally equal to $\sqrt{3}\sigma_\star/w$ above)
is a parameter that allows for uncertainty in the model,
and the last equality of \eqref{h} assumes that $\Ss/\zeta\gg\Sg$
and that stars dominate the total surface density of the disk $\St$,
which includes stars, gas and dark matter.%
\footnote{In practice, we only require that $\St\propto\Ss$, since $\zeta$ can be rescaled.}
A similar expression~-- but with $u^2$ replacing $w^2$~-- is motivated in \citet{Forbes+12}.%
\footnote{See also the expression for the total midplane pressure in \citealt{Elmegreen89},
which would lead to a similar expression for the scale height to that of \citet{Forbes+12}.}
Equation~\eqref{h} is a generalization of their formula that includes the thermal pressure,
$P\thm=\rho c\sound^2/\gamma\approx\tfrac{2}{3}\rho c\sound^2$ (if $\gamma=3/2$),
in addition to the turbulent pressure $P\turb=\tfrac{1}{3}\rho u^2$.%
\footnote{For example, naively comparing stellar velocity dispersion data \citep{Mogotsi+19} 
with gas velocity dispersion data \citep{Mogotsi+16}
for two different sets of galaxies suggests 
a mean ratio of stellar to gas 1D velocity dispersions in the range $5$--$10$.}

The equations of Sections~\ref{sec:magnetic_field}, \ref{sec:turbulence} and \ref{sec:observables}
can be used to calculate the vertically and azimuthally averaged 
turbulence and magnetic field properties at a radius $r$,
in terms of commonly observed quantities.

\begin{table}
\centering
  \def\arraystretch{1.1}
  \caption{Versions of the model for which scaling relations are derived.
  See Section~\ref{sec:alternative} for details.
  \label{tab:models}
} 
  \begin{tabular}{@{}ccccc@{}}
Model		&Driving scale		&Regime		&Correl.~time				&Mach no.		\\
\hline                                   		
S		&$\propto R\SN(t\sound)$	&a		&$\taue<\taur\Rightarrow\tau=\taue$	&$\Mach\ll A$		\\
		&			&b		&$\taue<\taur\Rightarrow\tau=\taue$	&$\Mach\gg A$		\\
		&			&c		&$\taur<\taue\Rightarrow\tau=\taur$	&$\Mach\ll A$		\\
		&			&d		&$\taur<\taue\Rightarrow\tau=\taur$	&$\Mach\gg A$		\\
\rule{0pt}{4ex}
Alt1		&$\propto h$		&		&$\tau=\taue$				&$\Mach=\const$		\\
\rule{0pt}{4ex}
Alt2 		&$\propto h$		&a		&$\tau=\taue$				&$\Mach\ll A$		\\
  		&			&b		&$\tau=\taue$				&$\Mach\gg A$		\\
    \hline
  \end{tabular}
\end{table}

\begin{table*}
\centering
  \def\arraystretch{1.050}
  \caption{Predicted scaling relations for parameters $h$, $l$, $u$ and $\tau$ for 
           our fiducial SN-driven turbulence model (Model~S)
           as well as two simpler alternative models Alt1 and Alt2.
           To derive scaling relations, 
           the models are subdivided into the asymptotic regimes $\Mach\ll A$ and $\Mach\gg A$
           as well as $\tau\eddy/\tau\renov<1$ and $\tau\eddy/\tau\renov>1$.
           See Section~\ref{sec:scaling} for details.
           \label{tab:scaling_int}
          }
  \begin{tabular}{@{}cllll@{}}
Model	&$h\propto$					&$l\propto$					&$u\propto$					&$\tau\propto$	\\
    \midrule
Sa	&$\St^{-1}T$	 				&$\Sg^{-0.37}\St^{-0.37}T^{0.21}$		&$\Sg^{-0.50}\St^{-0.16}\Sf^{1/3}T^{0.27}$	&$\Sg^{0.12}\St^{-0.21}\Sf^{-1/3}T^{-0.07}$	\\
Sb	&$\Sg^{-1.48}\St^{-1.49}\Sf^{0.99}T^{0.33}$	&$\Sg^{-0.92}\St^{-0.55}\Sf^{0.37}T^{-0.04}$	&$\Sg^{-0.74}\St^{-0.24}\Sf^{0.50}T^{0.17}$	&$\Sg^{-0.18}\St^{-0.31}\Sf^{-0.13}T^{-0.21}$	\\
Sc	&$\St^{-1}T$					&$\Sg^{-0.37}\St^{-0.37}T^{0.21}$		&$\Sg^{-0.50}\St^{-0.16}\Sf^{1/3}T^{0.27}$	&$\Sg^{19/17}\St^{2/17}\Sf^{-1}T^{13/34}$	\\
Sd	&$\Sg^{-1.48}\St^{-1.49}\Sf^{0.99}T^{0.33}$	&$\Sg^{-0.92}\St^{-0.55}\Sf^{0.37}T^{-0.04}$	&$\Sg^{-0.74}\St^{-0.24}\Sf^{0.50}T^{0.17}$	&$\Sg^{1.29}\St^{0.17}\Sf^{-1.12}T^{0.46}$	\\
\rule{0pt}{4ex}
Alt1	&$\St^{-1}T$					&$\St^{-1}T$					&$T^{1/2}$					&$\St^{-1}T^{1/2}$				\\
\rule{0pt}{4ex}
Alt2a	&$\St^{-1}T$					&$\St^{-1}T$					&$\St^{-1}\Sf^{1/3}T^{4/3}$			&$\Sf^{-1/3}T^{-1/3}$				\\
Alt2b	&$\St\Sf^{-2/3}T^{-2/3}$			&$\St\Sf^{-2/3}T^{-2/3}$			&$\St\Sf^{-1/3}T^{-1/3}$			&$\Sf^{-1/3}T^{-1/3}$				\\
    \bottomrule
  \end{tabular}
\end{table*}

\begin{table*}
\centering
  \def\arraystretch{1.050}
  \caption{Predicted scaling relations for the strengths of the isotropic ($b\iso$) and anisotropic ($b\ani$) components of the random magnetic field,
           and the strength ($\Bbar$) and pitch angle ($p_B$) of the mean magnetic field. 
           Expressions for $b\ani$ assume $q\Omega\tau/2\ll1$
           and those for $\Bbar$ assume $\min(1,C'_\alpha h/C_\alpha\tau u)\ge \Omega\tau$ 
           and $D\gg D\crit$.
           \label{tab:scaling_B}
          }
  \begin{tabular}{@{}cllll@{}}
Model	&$b\iso\propto$					&$b\ani\propto$							&$\Bbar\propto$  						&$\tan p_B\propto$						\\
    \midrule
Sa	&$\Sg^{0.003}\St^{0.34}\Sf^{1/3}T^{-0.23}$	&$q^{1/2}\Om^{1/2}\Sg^{0.07}\St^{0.23}\Sf^{1/6}T^{-0.26}$	&$q^{1/2}\Om\Sg^{0.13}\St^{0.13}T^{-0.29}$		&$q^{-1}\Om^{-1}\Sg^{-0.87}\St^{1.46}\Sf^{1/3}T^{-1.52}$	\\
Sb	&$\Sg^{1.24}\St^{0.74}\Sf^{-0.50}T^{0.33}$	&$q^{1/2}\Om^{1/2}\Sg^{1.15}\St^{0.59}\Sf^{-0.56}T^{0.23}$	&$q^{1/2}\Om\Sg^{0.32}\St^{0.19}\Sf^{-0.13}T^{-0.21}$	&$q^{-1}\Om^{-1}\Sg^{1.29}\St^{2.17}\Sf^{-1.12}T^{-0.54}$	\\
Sc	&$\Sg^{0.003}\St^{0.34}\Sf^{1/3}T^{-0.23}$	&$q^{1/2}\Om^{1/2}\Sg^{0.56}\St^{0.40}\Sf^{-1/6}T^{-0.03}$	&$q^{1/2}\Om\Sg^{0.13}\St^{0.13}T^{-0.29}$		&$q^{-1}\Om^{-1}\Sg^{0.12}\St^{1.79}\Sf^{-1/3}T^{-1.07}$	\\
Sd	&$\Sg^{1.24}\St^{0.74}\Sf^{-0.50}T^{0.33}$	&$q^{1/2}\Om^{1/2}\Sg^{1.88}\St^{0.83}\Sf^{-1.05}T^{0.57}$	&$q^{1/2}\Om\Sg^{0.32}\St^{0.19}\Sf^{-0.13}T^{-0.21}$	&$q^{-1}\Om^{-1}\Sg^{2.77}\St^{2.66}\Sf^{-2.11}T^{0.13}$	\\
\rule{0pt}{4ex}
Alt1	&$\Sg^{1/2}\St^{1/2}$				&$q^{1/2}\Om^{1/2}\Sg^{1/2}T^{1/4}$				&$q^{1/2}\Om\Sg^{1/2}\St^{-1/2}T^{1/2}$		     	&$q^{-1}\Om^{-1}\St T^{-1/2}$				\\
\rule{0pt}{4ex}
Alt2a	&$\Sg^{1/2}\St^{-1/2}\Sf^{1/3}T^{5/6}$		&$q^{1/2}\Om^{1/2}\Sg^{1/2}\St^{-1/2}\Sf^{1/6}T^{2/3}$		&$q^{1/2}\Om\Sg^{1/2}\St^{-1/2}T^{1/2}$		     	&$q^{-1}\Om^{-1}\Sf^{1/3}T^{1/3}$				\\
Alt2b	&$\Sg^{1/2}\St^{-1/2}\Sf^{1/3}T^{5/6}$		&$q^{1/2}\Om^{1/2}\Sg^{1/2}\St^{-1/2}\Sf^{1/6}T^{2/3}$		&$q^{1/2}\Om\Sg^{1/2}\St^{1/2}\Sf^{-1/3}T^{-1/3}$	&$q^{-1}\Om^{-1}\Sf^{1/3}T^{1/3}$				\\
    \bottomrule
  \end{tabular}
\end{table*}

\begin{table*}
\centering
  \def\arraystretch{1.050}
  \caption{As Table~\ref{tab:scaling_int}, but with the additional assumption that $\Sf\propto\Sg^{1.4}$.
           \label{tab:scaling_int_KS}
          }
  \begin{tabular}{@{}cllll@{}}
Model	&$h\propto$					&$l\propto$					&$u\propto$					&$\tau\propto$	\\
    \midrule
Sa	&$\St^{-1}T$	 				&$\St^{-0.37}\Sf^{-0.27}T^{0.21}$		&$\St^{-0.16}\Sf^{-0.021}T^{0.27}$		&$\St^{-0.21}\Sf^{-0.24}T^{-0.07}$		\\
Sb	&$\St^{-1.49}\Sf^{-0.064}T^{0.33}$		&$\St^{-0.55}\Sf^{-0.29}T^{-0.04}$		&$\St^{-0.24}\Sf^{-0.032}T^{0.17}$		&$\St^{-0.31}\Sf^{-0.26}T^{-0.21}$		\\
Sc	&$\St^{-1}T$					&$\St^{-0.37}\Sf^{-0.27}T^{0.21}$		&$\St^{-0.16}\Sf^{-0.021}T^{0.27}$		&$\St^{2/17}\Sf^{-0.20}T^{13/34}$		\\
Sd	&$\St^{-1.49}\Sf^{-0.064}T^{0.33}$		&$\St^{-0.55}\Sf^{-0.29}T^{-0.04}$		&$\St^{-0.24}\Sf^{-0.032}T^{0.17}$		&$\St^{0.17}\Sf^{-0.19}T^{0.46}$		\\
\rule{0pt}{4ex}
Alt1	&$\St^{-1}T$					&$\St^{-1}T$					&$T^{1/2}$					&$\St^{-1}T^{1/2}$				\\
\rule{0pt}{4ex}
Alt2a	&$\St^{-1}T$					&$\St^{-1}T$					&$\St^{-1}\Sf^{1/3}T^{4/3}$			&$\Sf^{-1/3}T^{-1/3}$				\\
Alt2b	&$\St\Sf^{-2/3}T^{-2/3}$			&$\St\Sf^{-2/3}T^{-2/3}$			&$\St\Sf^{-1/3}T^{-1/3}$			&$\Sf^{-1/3}T^{-1/3}$				\\
    \hline
  \end{tabular}
\end{table*}

\begin{table*}
\centering
  \def\arraystretch{1.050}
  \caption{As Table~\ref{tab:scaling_B}, but with the additional assumption that $\Sf\propto\Sg^{1.4}$.
           \label{tab:scaling_B_KS}
          }
  \begin{tabular}{@{}cllll@{}}
Model	&$b\iso\propto$				&$b\ani\propto$						&$\Bbar\propto$  				&$\tan p_B\propto$						\\
    \midrule
Sa	&$\St^{0.34}\Sf^{0.34}T^{-0.23}$	&$q^{1/2}\Om^{1/2}\St^{0.23}\Sf^{0.21}T^{-0.26}$	&$q^{1/2}\Om\St^{0.13}\Sf^{0.091}T^{-0.29}$	&$q^{-1}\Om^{-1}\St^{1.46}\Sf^{-0.29}T^{-1.52}$	\\
Sb	&$\St^{0.74}\Sf^{0.39}T^{0.33}$		&$q^{1/2}\Om^{1/2}\St^{0.59}\Sf^{0.26}T^{0.23}$		&$q^{1/2}\Om\St^{0.19}\Sf^{0.10}T^{-0.21}$	&$q^{-1}\Om^{-1}\St^{2.17}\Sf^{-0.19}T^{-0.54}$	\\
Sc	&$\St^{0.34}\Sf^{0.34}T^{-0.23}$	&$q^{1/2}\Om^{1/2}\St^{0.40}\Sf^{0.23}T^{-0.03}$	&$q^{1/2}\Om\St^{0.13}\Sf^{0.091}T^{-0.29}$	&$q^{-1}\Om^{-1}\St^{1.79}\Sf^{-0.24}T^{-1.07}$	\\
Sd	&$\St^{0.74}\Sf^{0.39}T^{0.33}$		&$q^{1/2}\Om^{1/2}\St^{0.83}\Sf^{0.29}T^{0.57}$		&$q^{1/2}\Om\St^{0.19}\Sf^{0.10}T^{-0.21}$	&$q^{-1}\Om^{-1}\St^{2.66}\Sf^{-0.13}T^{0.13}$	\\
\rule{0pt}{4ex}
Alt1	&$\St^{1/2}\Sf^{0.36}$			&$q^{1/2}\Om^{1/2}\Sf^{0.36}T^{1/4}$			&$q^{1/2}\Om\St^{-1/2}\Sf^{0.36}T^{1/2}$	&$q^{-1}\Om^{-1}\St T^{-1/2}$				\\
\rule{0pt}{4ex}
Alt2a	&$\St^{-1/2}\Sf^{0.69}T^{5/6}$		&$q^{1/2}\Om^{1/2}\St^{-1/2}\Sf^{0.52}T^{2/3}$		&$q^{1/2}\Om\St^{-1/2}\Sf^{0.36}T^{1/2}$	&$q^{-1}\Om^{-1}\Sf^{1/3}T^{1/3}$				\\
Alt2b	&$\St^{-1/2}\Sf^{0.69}T^{5/6}$		&$q^{1/2}\Om^{1/2}\St^{-1/2}\Sf^{0.52}T^{2/3}$		&$q^{1/2}\Om\St^{1/2}\Sf^{0.024}T^{-1/3}$	&$q^{-1}\Om^{-1}\Sf^{1/3}T^{1/3}$				\\
    \bottomrule
  \end{tabular}
\end{table*}

\section{Scaling relations}
\label{sec:scaling}
We now derive scaling relations for key turbulence and magnetic quantities.
These can be obtained using straightforward algebra,
but we have provided the link to a tool which facilitates this in Section~\ref{sec:model}.
We focus on the power law exponents,
but a sample calculation 
of the proportionality coefficients 
is provided in Sec.~\ref{sec:coef}.
Adjustable parameters $\zeta$, $\psi$, $C_\alpha$, $K$ and $\beta$ 
are assumed to be constant or to vary weakly between galaxies,
so the exponents do not depend on them.
We set $A_1=A_2\equiv A$ for simplicity
and restrict our analysis to the asymptotic cases $\Mach\ll A$ and $\Mach\gg A$, 
so that $w$ can be replaced, respectively, by $c\sound$ or $u$ in equation~\eqref{h},
but we comment on the regime $\Mach\approx A$ in Section~\ref{sec:transonic}.

To obtain scaling relations for $b\ani$ and $\mean{B}$, additional approximations are necessary.
In equation~\eqref{b_ani_noUz} for $b\ani$, we assume that $q\Omega\tau/2 \ll 1$.
This condition is likely satisfied in the Solar Neighbourhood, for example,
where $q\Omega\tau\approx0.14$ assuming $\Omega=27.5\kmskpc$, $q=1$ and $\tau=5\Myr$ \citep{Hollins+17,Chamandy+Shukurov20}.
Equation~\eqref{b_ani_noUz} then simplifies to
\begin{equation}
  b\ani= \left(\frac{2}{3}q\Omega\tau\right)^{1/2}b\iso,
\end{equation}
where $b\iso$ is still given by equation~\eqref{b_iso}.

In order to be able to write down a scaling relation for $\mean{B}$ ,
we assume $\min\left(1,C'_\alpha h/C_\alpha\tau u\right)\ge \Omega\tau$,
so that, from equation~\eqref{alphak_approx} we have the standard result 
\citep{Krause+Radler80,Ruzmaikin+88}
\begin{equation}
  \label{alphak_Krause}
  \alpha\kin= \frac{C_\alpha\tau^2u^2\Omega}{h}.
\end{equation}
In addition, we assume that the dynamo is highly supercritical, i.e.~$D/D\crit\gg1$. 
This assumption is generally satisfied in the inner parts of galaxies. 
With these assumptions, equation~\eqref{Bbar_noUz} simplifies to
\begin{equation}\label{Bbar_Krause}
  \Bbar=  K \,\frac{\uppi l}{h}\left(\frac{D R_\kappa}{D\crit}\right)^{1/2}B\eq,
\end{equation}
with
\begin{equation}\label{Dk_Krause}
  D = -\frac{9C_\alpha q h^2\Omega^2}{u^2},
\end{equation}
where we have made use of equations~\eqref{Dk}, \eqref{Reynolds}, \eqref{eta} and \eqref{alphak_Krause}.
Note that $D\propto q\Omega^2$, 
so there is some tension between assuming both $D\gg D\crit$ and $q\Omega\tau/2\ll1$. 
Both assumptions may be satisfied simultaneously for small $\tau$.
Upon substituting equations~\eqref{Dk_Krause} and \eqref{Beq} in equation~\eqref{Bbar_Krause},
we find that $\Bbar\propto l\Omega (C_\alpha q\rho R_\kappa)^{1/2}$.
Simulations suggest $R_\kappa\approx0.3$ \citep{Mitra+10},
while a recent analytic calculation gives $R_\kappa=21(1+\tilde{\xi})/27$ with $\tilde{\xi}=b^2/B\eq^2$ 
\citep{Gopalakrishnan+Subramanian23}.
In the present work, $R_\kappa$ is assumed to be constant.

We now have all of the expressions needed to obtain scaling relations for 
$h(r)$, $u(r)$, $l(r)$, $\tau(r)$, $b\iso(r)$, $b\ani(r)$, $\Bbar(r)$ and $p_B(r)$
in terms of the observables $\Sigma_\star(r)$ or $\St(r)$, $\Sigma(r)$, $\Sigma\sfr(r)$, 
$\Omega(r)$, $q(r)$ and $T(r)$.
However, the scaling relations will depend 
on whether $\taue>\taur$ or $\taue<\taur$ (equation~\ref{tau}),
and on whether $u\ll A c\sound$ or $u\gg A c\sound$ (equation~\ref{w}).
For transonic turbulence we might expect scaling relation exponents in between those
for the subsonic and supersonic cases.

\subsection{Alternative turbulence models}\label{sec:alternative}
Turbulence driving in galaxies is still not very well-understood 
and there is still a lack of consensus, e.g., about the driving scale(s).
Thus, we choose to explore the implications of making different assumptions about turbulence driving,
as summarized in Table~\ref{tab:models}.
Model~S is our fiducial model 
and uses the framework summarized above for turbulence driven by expanding isolated SNRs.
We include two simpler models for comparison. 
Model~Alt1 is our minimalistic model,  which assumes $l\propto h$,
$u\propto c\sound$ and $\tau\propto l/u$
(in deriving the scaling relation exponents, 
equalities can be replaced with proportionalities).
Model~Alt2 is our hybrid model and assumes $l\SN\propto h$
and $\tau\propto l/u$, 
but retains equation~\eqref{u_noSBs} for $u$, 
which permits two asymptotic regimes ($\Mach\ll A$ and $\Mach\gg A$), as in Model~S.
The \textsc{python} tool mentioned in Section~\ref{sec:model}
can be used to explore a wider set of combinations of assumptions about turbulence.

\subsection{Expressions}\label{sec:expressions}
Scaling relations for Models~S, Alt1 
and Alt2 are given in Tables~\ref{tab:scaling_int} and \ref{tab:scaling_B}.
Where reduced fractional forms of exponents are cumbersome to write, we have rounded to decimal values.
We choose to write the scaling relations in terms of $\St$ rather than $\Ss$,
but we refer to them somewhat interchangeably below.
Note that the dependency on $\zeta$ is the same as the dependency on $1/\St$.

\subsection{Relations between observables}\label{sec:relations}
An important caveat is that the observables $q(r)$, $\Om(r)$, $\Sg(r)$, $\Sf(r)$,
$\Sigma_\star(r)$ and $T(r)$ 
are not mutually statistically independent. 
For example, the SFR and gas surface densities 
are typically related by the Kennicutt-Schmidt (KS) law,
\begin{equation}\label{KS}
  \Sf \propto \Sg^N,
\end{equation}
with $N\approx1.4$ \citep{Kennicutt+Evans12}.
As this relation is rather tight and universal,
we have used the KS law to replace $\Sg$ by $\Sf^{1/N}$ in Tables~\ref{tab:scaling_int_KS} and \ref{tab:scaling_B_KS}, 
eliminating the explicit dependence on $\Sg$.
We focus on Tables~\ref{tab:scaling_int_KS} and \ref{tab:scaling_B_KS} in the discussion below.

Other correlations between our observables have also been measured,
like the so-called spatially resolved star-forming main sequence
relating $\Sf$ and $\Sigma_\star$ \citep[e.g.][]{Cano-diaz+16,Baker+22}.
However, this relation shows quite a lot of scatter.
Moreover, the power law exponent, $\del\ln\Sigma_\star/\del\ln\Sf$, is not well-constrained, 
and may be sensitive to resolution \citep{Hani+20}.
For this reason, we treat $\St\approx\Sigma_\star$ and $\Sf$ as mutually independent quantities. 

\subsection{Consistency of models across regimes}\label{sec:transonic}
In Model~S (SN-driven model), the exponents in Tables~\ref{tab:scaling_int_KS} and \ref{tab:scaling_B_KS}
are remarkably consistent when transitioning between the four different physical regimes.
Thus, the results for Model~S are fairly insensitive to variations in the Mach number and the ratio of the eddy turnover and renovation times.
By contrast, in Model~Alt2 (hybrid model) the rms turbulent velocity $u$ has very different power law exponents for the two limiting cases 
($\Mach\ll A$ and $\Mach\gg A$).
This implies a sharp transition in the transonic regime, 
which seems unlikely from a physical standpoint,
lending support to the claim that Model~S is the most realistic among the models. 

\subsection{Comparison with observations}\label{sec:comparison_observations}
We now compare our results with previously published observations, 
continuing to focus on Tables~\ref{tab:scaling_int_KS} and \ref{tab:scaling_B_KS}, 
which take into account the KS law, i.e. the empirical relation between $\Sg$ and $\Sf$ (equation~\ref{KS}).

\subsubsection{Turbulent velocity}\label{sec:turbulence_disk}
As seen in Table~\ref{tab:scaling_int_KS}, 
Model~S predicts that the rms turbulent velocity $u$ depends very weakly on $\Sf$ and depends weakly on $\St$ and $T$.
One might then expect the variation in the velocity dispersion within a galaxy or between galaxies to be weak.
This is borne out in the data. 
Based on second moment maps of THINGS galaxies \citep{Walter+08}, 
\citet{Leroy+08} find the 1D velocity dispersion for THINGS galaxies to be $11\pm3\kms$.
The gas velocity dispersion within a given galaxy 
is also fairly constant \citep[e.g.][]{Mogotsi+16,Chemin+16}.
Thus, the predicted weak dependence of $u$ on $\St$, $\Sf$ and $T$ 
is qualitatively consistent with observations \textit{at low redshift}.

From Table~\ref{tab:scaling_int_KS}, 
Model~Alt1 predicts $u$ to be independent of the other observables except $T$, 
where the $T^{1/2}$ dependence is stronger than for Model~S.
By contrast, Model~Alt2 predicts a strong dependence on the various observables
as well as different signs of the exponents for the subsonic and supersonic cases,
and thus disagrees with observations.

\subsubsection{Relationship between field strength and $\Sf$}\label{sec:j_disk}
It has been observed in many galaxies that the total magnetic field strength correlates with $\Sf$,
and many authors have obtained values for $j$ in the relation $B\propto\Sf^j$.
\citet{Niklas+Beck97} find $j=0.34\pm0.08$, assuming energy equipartition between magnetic field and cosmic rays.
\citet{Chyzy08} finds a tight correlation looking at the spatial variation of $B$ and $\Sf$ in the Virgo Cluster spiral galaxy NGC~4254,
and obtains $j=0.18\pm0.01$. 
They also find a tight correlation between the random component 
(calculated from the unpolarized emission)
and $\Sf$, with a power law exponent of $0.26\pm0.01$.
\citet{Chyzy+11} find $j=0.30\pm0.04$ for dwarf irregular galaxies.
\citet{Heesen+14} infer $j=0.30\pm0.02$ for $17$ THINGS galaxies.
\citet{Vaneck+15} find $j=0.19\pm0.03$, using results that assume energy equipartition.
\citet{Wang+Lilly22} interpret \citet{Heesen+14} and other observations and assume energy equipartition to obtain $j=0.15\pm0.06$.
They show how an exponential profile for $\Sf(r)$~-- seen in many disk galaxies~-- 
can be explained by their model as long as $0.1\lesssim j\lesssim0.2$.

For the galaxy NGC~6946, \citet{Tabatabaei+13a} obtain $j=0.14\pm0.01$,
and, if $B$ is replaced by the 
turbulent component (calculated from the unpolarized emission),
they obtain an equally strong correlation with a slightly different exponent of $0.16\pm0.01$.
\citet{Vollmer+20} obtain $j$ values between $0.03$ and $0.92$, 
with a lot of scatter and a best fit value of either $0.13\pm0.09$ or $0.44$, depending on the statistical technique adopted.
There has also been a flurry of very recent work. 
\citet{Heesen+23} find $j=0.182\pm0.004$, but with considerable scatter, 
or $j=0.22$ when they use the mean values of each of the 12 galaxies considered.
However, when they account for cosmic-ray transport, they infer a value of $j=0.284\pm0.006$ instead of $0.182\pm0.004$.
These authors also report that $B$ is found to be more tightly correlated 
with the gas surface density $\Sigma= \Sigma_\mathrm{HI}+\Sigma_\mathrm{H_2}$ than with $\Sf$.
\citet{Tabatabaei+22} study the galaxy M33 and find bi-modal behaviour. 
For the high-SFR regime they obtain $j=0.25$--$0.26$ and for the low-SFR regime they obtain $j=0.10$--$0.12$.
\citet{Basu+17} obtains $j=0.35\pm0.03$ for the dwarf irregular galaxy IC~10.
Finally, \citet{Manna+Roy23} obtain $j=0.31\pm0.06$ for a sample of seven galaxies, using the equipartition assumption.

In the present work we have derived expressions for 
$b\iso$, $b\ani$ and $\mean{B}$,
and scaling relations are presented in Table~\ref{tab:scaling_B_KS}.
Observers generally infer that the 
isotropic random
component of the magnetic field 
dominates in most galaxies.
Moreover, when a mean component is modeled from the observations, 
this component is usually determined to be quite weak,
suggesting that most of the polarized emission 
is not due to the mean field, 
but to an anisotropic small-scale component \citep{Beck+19}.
At face value, this would imply that, typically, $b\iso>b\ani>\mean{B}$. 
However, 
the mean field may often be underestimated if, as seems likely, 
it contains significant power at scales that cannot be resolved.

Examining the scaling relations for $b\iso$, $b\ani$ and $\mean{B}$ 
in Table~\ref{tab:scaling_B_KS} for Model~S,
we see that they are related to $\Sf$ by power laws with exponents 
in the ranges $0.34$--$0.39$, $0.21$--$0.29$ and $0.09$--$0.10$, respectively.
Thus, if $\bfb\iso$ is the dominant field component, a suitable weighted average might give $j\sim0.30$, 
whereas if $\meanv{B}$ is the dominant component, we might expect $j\sim0.15$.
While a detailed comparison between theory and observation is clearly not possible here,
the theoretical and empirical
estimates for $j$
agree rather well.
Furthermore, the prediction of our model that $j$ is higher for the isotropic 
random component than for the mean component aligns with the findings of \citet{Chyzy08}
and \citet{Tabatabaei+13a} (though the latter work finds only a marginal difference),
as well as with those of \citet{Tabatabaei+22} if the mean field is relatively stronger in the low-SFR regions.

On the other hand, Model~Alt1 predicts $j=0.36$, 
which is reasonably consistent with observations,
but Model~Alt2 predicts an exponent $0.69$ for $b\iso$, $0.52$ for $b\ani$
and a range $0.02$--$0.36$ for $\mean{B}$,
which are too high to explain observations.

\begin{figure*}
    \centering
    \includegraphics[ width = 0.48\textwidth, clip=true, trim= 11 0 5 0]{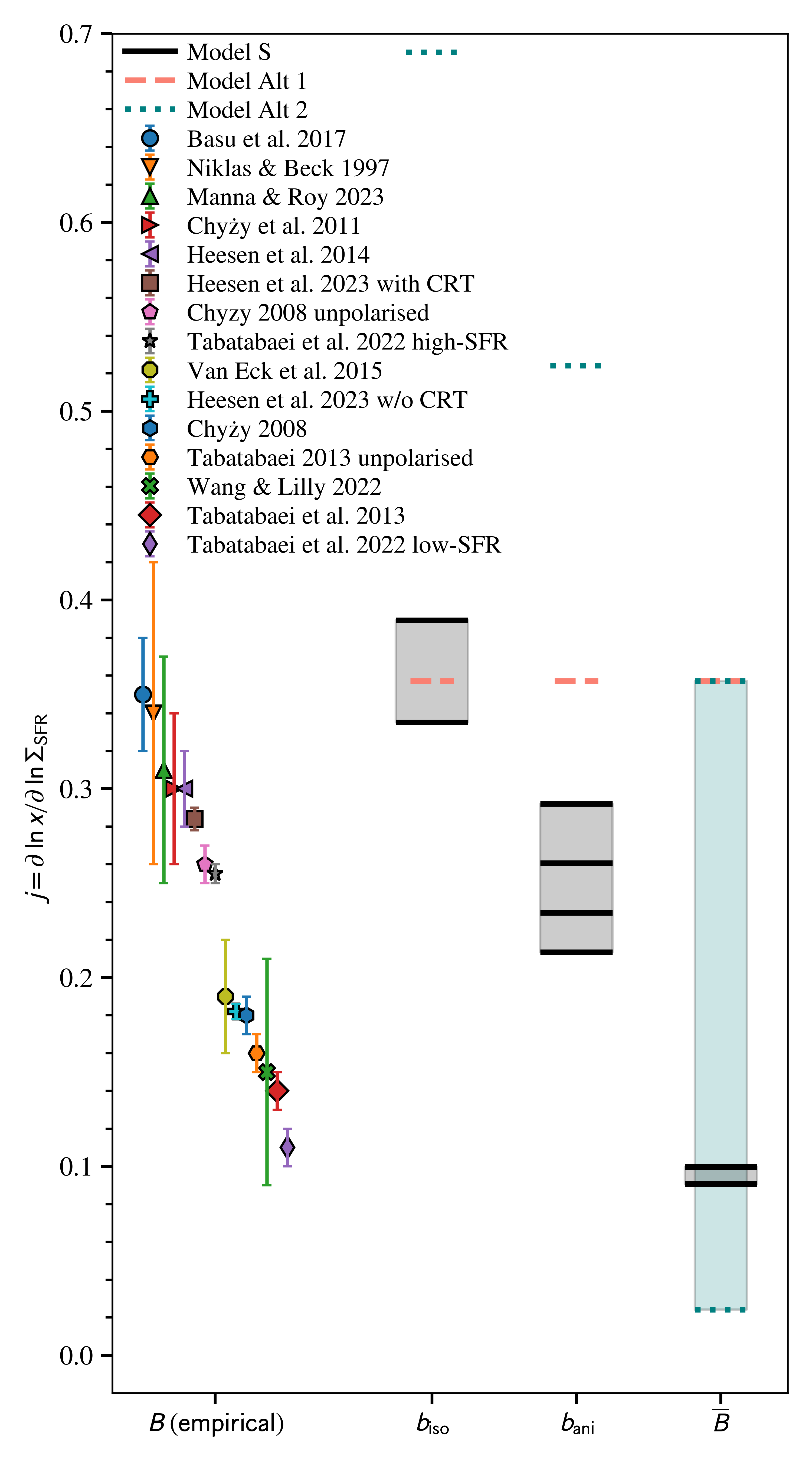}
    \includegraphics[width = 0.48\textwidth, clip=true, trim= 11 0 5 0]{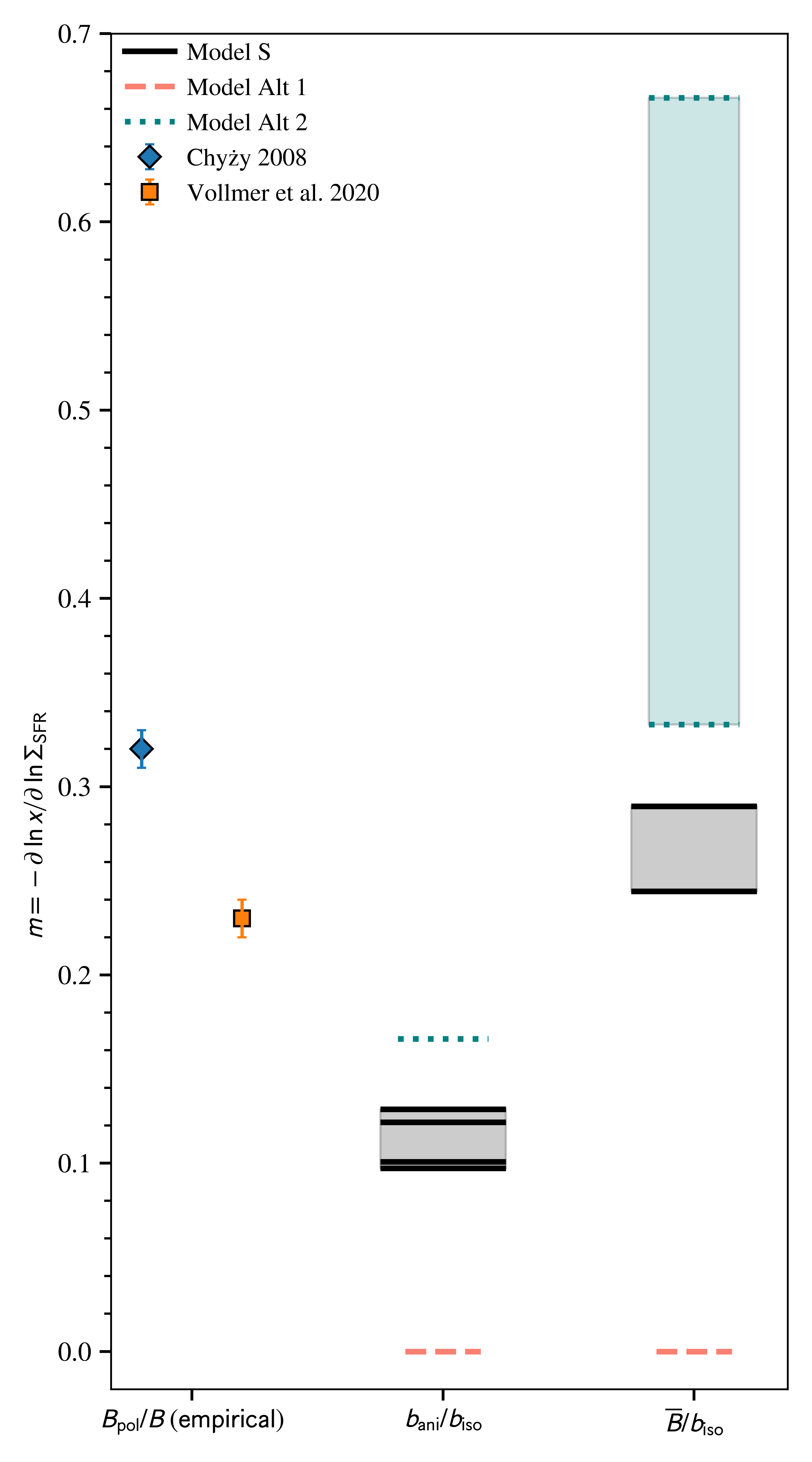}
    \caption{The left panel compares the quantity $j=\del\ln B/\del\ln\Sf$ inferred from observations (symbols) with the values of $\del\ln b\iso/\del\ln\Sf$, $\del\ln b\ani/\del\ln\Sf$, and $\del\ln\Bbar/\del\ln\Sf$ obtained from Models~S, Alt1, and Alt2 (horizontal lines). The range spanned by the data is quite consistent with the predictions of the fiducial model, Model~S.
    In the right panel, 
    data points show the scaling relation exponent 
    $m=-\del\ln(B\pol/B)/\del\ln\Sf$ 
    inferred from observations.
    Horizontal lines and ranges show the predictions of our models when $B\pol$ is assumed to be dominated by $b\ani$ or $\mean{B}$, and $B$ by $b\iso$.
    Predictions of the fiducial model, Model~S, 
    with parameters chosen such that $\mean{B}\gg b\ani$, 
    are quite consistent with the data.}
    \label{fig:B_Sf}
\end{figure*}

These results are summarized in the left panel of Figure~\ref{fig:B_Sf},
where it can be seen that the range predicted by Model~S
corresponds reasonably well to the range spanned by the data.
Keep in mind that in our model $j$ would be a weighted average 
of the $j$-values for $b\iso$, $b\ani$ and $\Bbar$,
so the observed values of $j$ would be expected to lie somewhere between 
the highest and lowest predictions.
Taking our Model~S at face value,
the upper cluster of points between $0.25$ and $0.35$ is just what might be expected
if $b\iso$ dominates but $b\ani$ and/or $\mean{B}$ are significant,
whereas the lower cluster of points between about $0.1$ and $0.2$ 
could arise if the mean magnetic field is particularly strong.

A theoretical model by \citet{Schleicher+Beck13} 
(see also \citealt{Schleicher+Beck16}) 
predicts $B\propto 
\Sf^{1/3}$,
whereas the model by \citet{Schober+23} (see also \citealt{Schober+16})
predicts $B\propto\Sf^{1/3}\Sg^{1/6}$,
which gives $B\propto\Sf^{0.45}$ if $\Sf\propto\Sg^{1.4}$ is assumed.
Those models have similarities with one another.
They are also somewhat similar to Model~Alt2a of the present work,
but their prescriptions for the gas scale height are different from ours.

\subsubsection{Relationship between field strength and density}
\citet{Manna+Roy23} obtain $B\propto\rho^{0.40\pm0.09}$ for a sample of seven galaxies.
Using equations \eqref{b_iso}, \eqref{Beq}, \eqref{u_noSBs} and \eqref{l_SN}, 
we find that $b\iso$ is almost independent of $\rho$ for Models~Sa and Sc,
but for Models~Sb and Sd we obtain a power law exponent of $1/2$
since $u$ cancels out in equation~\eqref{b_iso} with $B\eq$ given by equation~\eqref{Beq}.
For transonic turbulence, we would therefore expect a positive exponent somewhere in between $0$ and $1/2$,
so Model~S is broadly consistent with observations.
For Model~Alt1, $b\iso\propto\rho^{1/2}$ since $u$ is independent of $\rho$.
For Model~Alt2a, $l\SN\propto h$, which is independent of $\rho$, 
so $b\iso\propto\rho^{1/2}$, 
while for Model~Alt2b, $u$ cancels so that again $b\iso\propto\rho^{1/2}$. 
Thus, predictions of Models~S, Alt1 and Alt2 all broadly agree with the empirical relation.

\subsubsection{Relationship between field alignment and $\Sf$}\label{sec:ordering}
In the study of NGC~4254 mentioned above, \citet{Chyzy08} also found 
that the ratio of the strength of the magnetic field inferred from polarized emission
(comprising mean and anisotropic turbulent components) to that inferred from total emission
is negatively correlated with $\Sf$, i.e.~$B\pol/B\propto \Sf^{-m}$, with $m=0.32\pm0.01$.
More recently, \citet{Vollmer+20} measured the same relation, but using data from several galaxies.
They find a tight correlation, with $m=0.22$ or $0.24$, depending on the fitting method.

To make a comparison with the model, 
let us first assume that the polarized emission is dominated by $b\ani$ and the total emission by $b\iso$
and consider the ratio $b\ani/b\iso$ using results from Table~\ref{tab:scaling_B_KS}. 
For Model~S, we obtain $m=0.13$, $0.13$, $0.11$ and $0.10$, respectively.
Therefore, the correct sign of the exponent is reproduced in the model, 
but its magnitude is lower than the observational values.
For Model~Alt1, we obtain $m=0$,
which is in disagreement with the observations.
For Model~Alt2, we obtain $m=0.17$, 
which is quite close to the observational values.

However, if $\mean{B}$ actually dominates the polarized emission, 
then we should use $\mean{B}/b\iso$ instead of $b\ani/b\iso$.
In this case, Model~S predicts values of $m$ in the range $0.25$-$0.29$, 
which is very close to the observed values.
On the other hand, Model~Alt1 again predicts $m=0$, 
while Model~Alt2a and Alt2b predict $m=0.33$ and $m=0.67$, respectively.
These results are summarized 
in the right panel of Figure~\ref{fig:B_Sf},
which illustrates that Model~S is quite consistent with the data
only if $B\pol/B\sim \mean{B}/b\iso$, which implies $b\iso>\mean{B}> b\ani$.
This ordering of $b\iso$, $\mean{B}$ and $b\ani$ is possible in our model, 
given the freedom in the parameter choices.
However, if this interpretation is correct, it would suggest that the mean field
may often be underestimated when interpreting radio observations of nearby galaxies,
since such studies typically find that $b\ani>\mean{B}$.
We note, however that different kinds of averages are used in theory and observations,
which complicates their comparison \citep{Beck+19}.

\subsubsection{Other dependencies of magnetic field properties}\label{sec:magnetic_disk}
\citet{Vaneck+15} explored the dependencies of the total field strength, 
mean field strength and mean field pitch angle, 
as obtained from observations 
(supplemented by the usual assumption of equipartition between magnetic and cosmic ray energies).
The observables they explored as independent quantities were the \HI surface density $\Sigma_I$, 
molecular gas surface density $\Sigma_2$, 
their sum $\Sigma_I+\Sigma_2$, $\Sf$, $\Omega$, $q\Omega$ and $q\Omega^2$.
They also looked for statistical relationships between $B$, $\mean{B}$ and $p_B$.%
\footnote{What \citet{Vaneck+15} interpret as $\Bbar$ may in some cases contain a contribution
from anisotropic random field ($b\ani$ in our model; c.~f.~\citealt{Beck+19}). 
For the purpose of comparing our model results with the findings of \citet{Vaneck+15},
we ignore this complication and directly compare our model with their final results; a re-analysis of the data is left for future work.}
Of these 24 pairs of variables, $10$ were found to exhibit correlations.
We now look at each of these and compare them with the findings in Table~\ref{tab:scaling_B_KS}.

As mentioned in Section~\ref{sec:j_disk}, \citet{Vaneck+15} obtained $B\propto\Sf^{0.19\pm0.03}$, 
with $B$ the strength of the total field.
They also obtained $B\propto\Sigma_2^{0.21\pm0.04}$ and $B\propto(\Sigma_I+\Sigma_2)^{0.24\pm0.07}$.
These latter two can be interpreted as a confirmation of Schmidt/KS laws, 
given the relationship found between $B$ and $\Sf$.
However, the value of $N$ implied is slightly lower than $1.4$.

The fourth and fifth correlations obtained were between $B$ and $q\Omega$ (power law not given) 
and $B\propto(q\Omega^2)^{0.14\pm0.04}$.
If the total field strength were dominated by $b\iso$, 
then Table~\ref{tab:scaling_B_KS} would predict no dependence, 
i.e.~no correlation. 
If dominated by $\mean{B}$, then Table~\ref{tab:scaling_B_KS} would predict a power of $0.5$. 
Thus, the observed value of $0.14\pm0.04$ is intermediate between these values,
suggesting that the mean field tends to be significant, but not dominant. 
The dependence of $b\ani$ on $q$ and $\Omega$ 
(but not explicitly on $q\Omega^2$ in the model) likely also plays a role.

The sixth, seventh and eighth correlations found by \citet{Vaneck+15} 
were $\tan p_B \propto\Sigma_2^{-0.10\pm0.08}$, $\tan p_B \propto(\Sigma_I+\Sigma_2)^{-0.25\pm0.13}$
and $\tan p_B\propto\Sf^{-0.15\pm0.09}$.
These are approximately related by the Schmidt/KS law, as expected, 
so let us consider only the last relation.
The exponent is similar to that predicted for Model~S in Table~\ref{tab:scaling_B_KS},
where we obtain $-(0.13$--$0.29)$, 
depending on the exact regime ($\tau\eddy/\tau\renov<1$ or $>1$, and $\Mach\ll1$ or $\gg1$).
On the other hand, Model~Alt1 predicts no dependence
and Model~Alt2 predicts an exponent equal to positive $1/3$,
so Model~S is compatible with observations, but Models~Alt1 and Alt2 are not.

\subsubsection{Relation between mean and total field strengths}\label{sec:Bbar-B}
The ninth correlation found by \citet{Vaneck+15} 
can be described by the power law $\mean{B}\propto B^{0.76\pm0.23}$.
Given that the powers of the observables generally have the same sign in Model~S 
for $b\iso$, $b\ani$ and $\mean{B}$ (Table~\ref{tab:scaling_B_KS}),
one would expect the scaling between $\mean{B}$ and $B$ to have a positive exponent, consistent with the above observation. 
The exception is the power of the gas temperature $T$.
However, $T$ does not usually vary as much as some of the other parameters in galaxies,
where warm gas is the dominant volume-filling phase of the ISM, 
and retains a temperature of order $10^4\K$.
In addition, one might expect the power to be $<1$, as observed, 
since the exponents in the scaling relation for $\mean{B}$ are generally smaller than those in the scaling relations for $b\iso$ and $b\ani$, 
but not very small because $\mean{B}$ itself contributes significantly to $B$.
Therefore, Model~S again gives results that are broadly consistent with observations.
For Models~Alt1 and Alt2,
the exponents of $\St$ tend to change sign when going from $\mean{B}$ to $b\iso$,
so the correlation between $\mean{B}$ and $B$ seems harder to explain 
using these alternative models.

\subsubsection{Relation between mean field strength and pitch angle}\label{sec:Bbar-pB}
The final correlation detected by \citet{Vaneck+15} -- between $\mean{B}$ 
and $p_B$ -- was the most statistically significant.
Though the power law fit was not provided,
the magnitude of $p_B$ was found to decrease as $\mean{B}$ increases.
This trend was confirmed by \citet{Beck+19}, using updated data.
Given this negative correlation, 
one might expect the exponents in the scaling relations for $\mean{B}$ and $p_B$ 
to generally have opposite signs.
The powers of $q$, $\Omega$ and $\Sf$ have opposite signs 
in the expressions for $\mean{B}$ and $\tan p_B$ of Model~S,
while those of $\St$ have the same sign, and those of $T$ are the same in three of four cases.
Thus, more work is needed to understand this empirical relation using dynamo models.

\subsubsection{Cases where no empirical scaling relation was found}\label{sec:null}
For the remaining $14$ relationships explored by \citet{Vaneck+15}, 
no statistically significant correlation was found.
Can these null results be explained by Model~S?
One such result was that $\mean{B}$ was not found to depend significantly on any of 
$\Sf$, $\Sigma_I$, $\Sigma_2$, $\Sigma_I+\Sigma_2$, $\Omega$, $q\Omega$ or $q\Omega^2$.
Model~S predicts only a weak dependence on most of these, 
but one would expect to find $\mean{B}\propto (q\Omega^2)^{1/2}$ in all of the models.
Moreover, $\tan p_B$ would be expected to scale as $(q\Omega)^{-1}$ in all the models,
whereas no such correlation was found by \citet{Vaneck+15}.
More work is needed to understand why models and data seem to disagree about this particular aspect.

\subsubsection{Arm-interarm contrast of random and mean fields}\label{sec:arm-interarm}
Model~S predicts (Table~\ref{tab:scaling_B_KS}) that for \textit{lower} $\Sf$ or $\St$, 
the ratio $\mean{B}/b\iso$ should increase.
This suggests that the field should be coherent on larger scales in the interarm regions,
where $\Sf$ and $\St$ are smaller than in the arms.
When the arm-interarm contrast of $\mean{B}/b\iso$ can be estimated from observations
this ratio is inferred~--
particularly for the galaxy NGC~6946 where the data is of good quality~-- to 
be higher within the interarm regions \citep{Beck+Hoernes96},
consistent with our model predictions.
On the other hand, Model~Alt1 predicts that $\mean{B}/b\iso$ 
should decrease with $\St$ but not depend on $\Sf$
and Model~Alt2 predicts that $\mean{B}/b\iso$ should decrease with $\Sf$
but not depend on (increase with) $\Ss$ if $\Ma\ll A$ ($\gg A$),
so again, Models~Alt1 and Alt2 do not agree with the observations 
as well as our SN-driven turbulence model, Model~S.

\subsection{Comparison with simulations}\label{sec:comparison_simulations}
\citet{Elstner+Gressel12} is the only work of which we are aware 
that obtains such scaling relations from direct numerical simulations \citep{Gressel+08b,Gressel+11}. 
These authors employ simulations of SN-driven turbulence to study dynamo action
in a small section of a galaxy.
They provide scaling relations for a few quantities, based on nine simulations.
In seven of their runs, the final magnetic field energy is of order the turbulent energy,
but in the other two the simulation ends when $B\ll B\eq$, so those runs are less relevant.
They also find that both vertical outflows and diamagnetic pumping are important,
but these effects have been neglected in deriving the scaling relations found in the present work.

Given the differences between their model and ours, 
it is perhaps most useful to compare results for the turbulent diffusivity $\eta$
because this quantity is not expected to be sensitive to details such as the dynamo saturation mechanism \citep{Brandenburg+Subramanian05a}.
\citet{Elstner+Gressel12} find $\eta\propto\sigma^{0.4}n^{0.4}\Omega^{-0.55}$,
where $\sigma$ is the SFR density in dimensions of $[\text{T}]^{-1}[\text{L}]^{-2}$,
and, as above, $n$ and $\Omega$ are the number density of gas and the rotation angular speed, 
respectively.
The value of $\sigma$ is determined from their input SN rate 
assuming a constant initial mass function, 
so we can replace $\sigma$ with $\Sf$, 
resulting in $\eta\propto\Sf^{0.4}n^{0.4}\Omega^{-0.55}$.

To compare this result with our models, we derive scaling relations using equation~\eqref{eta},
the scaling relations of Table~\ref{tab:scaling_int}, and equation~\eqref{rho}, and assume $\rho/n=\const$.
We use Table~\ref{tab:scaling_int} rather than Table~\ref{tab:scaling_int_KS} because
the overall SN rate density is varied as a parameter in the \citet{Elstner+Gressel12} simulations.%
\footnote{In their simulations, 
the SN rate density is higher where the gas density is higher,
but the overall normalization of the SN rate surface density is set by hand,
independently of the gas profile,
and varied from one simulation to another 
to isolate the effects of this one parameter \citep{Bendre+15}.}
Neglecting the temperature dependence, 
Models~Sa-d respectively give $\eta\propto\Sf^{1/3}n^{-0.9}\St^{0.4}$, $\eta\propto\Sf^{0.2}n^{-0.7}\St^{0.2}$,
$\eta\propto\Sf^{-1/3}n^{0.1}\St^{-0.3}$ and $\eta\propto n^{-0.1}\St^{-0.4}$ 
(where in the last relation we have neglected the very weak dependence on $\Sf$).
Models~Alt1, Alt2a and Alt2b give $\eta\propto\St^{-1}$, $\eta\propto\Sf^{1/3}\St^{-2}$ and $\eta\propto\Sf^{-1}\St^{2}$, respectively.
Given that $\Omega$ depends on the gravitational potential, 
$\Omega$ and $\St$ are not completely independent from one another.
Thus, the absence of $\Omega$ in our scaling relations for $\eta$ 
does not necessarily imply disagreement with \citet{Elstner+Gressel12}.
For this reason, let us focus on $\Sf$ and $n$.
In Models~Sa, Sb and Alt2a, we find $\Sf$ exponents of $1/3$, $0.2$ and $1/3$, respectively, 
which are close to the value of $0.4$ obtained by \citet{Elstner+Gressel12}.
In the other models, the agreement is poorer, with exponents $-1/3$, $\approx0$, $0$ and $-1$,  
for Models~Sc, Sd, Alt1 and Alt2b, respectively.
Now turning to the dependence on $n$, we obtain a positive exponent only for Model~Sc ($0.1$), 
whereas other models give exponents ranging from $-0.9$ to $0$, 
which is in disagreement with the value of $0.4$ obtained by \citet{Elstner+Gressel12}.
We can conclude that the level of agreement between our predicted scaling relation 
for $\eta$ and that found by \citet{Elstner+Gressel12} is rather poor. 
However, the dependence on $\Sf$ agrees quite well in Model~S whenever $\taue<\taur$
and in Model~Alt2a, where $\Mach\ll A$
(although \citealt{Gressel+08a} report mildly supersonic turbulence 
for the simulations of that paper, so Alt2b is probably more relevant than Alt2a, 
but there the model predicts a dependence on $\Sf$ different from that seen in the simulations). 

In addition, \citet{Gressel+08b} report that the ratio of the strength 
of the mean field to that of the random field 
is related to the SN rate (SFR density) as $\mean{B}/b\propto\Sf^{-0.38\pm0.01}$,
while for similar simulations by \citet{Bendre+15} the authors report $\mean{B}/b\propto\Sf^{-0.30\pm0.07}$.
Generally, we expect $b\iso>b\ani$ \citep{Beck+19}, 
so we focus on the ratio $\mean{B}/b\iso$, 
using Table~\ref{tab:scaling_B}.
For Model~Alt1 we find no dependence of $\mean{B}/b\iso$ on $\Sf$.
For Models~Alt2a and Alt2b we find $\mean{B}/b\iso\propto\Sf^{-1/3}$ and $\mean{B}/b\iso\propto\Sf^{-2/3}$, respectively.
For Models~Sa and Sc, we find $\mean{B}/b\iso\propto\Sf^{-1/3}$,
and for Models~Sb and Sd we find $\mean{B}/b\iso\propto\Sf^{0.37}$.
Thus, for Models~Alt2a, Sa and Sc (all of which have $\Mach\ll A$) the agreement is good, 
for Model~Alt2b the agreement is fair, and for Models~Sb and Sd the agreement is poor.
Differences may be partly attributable 
to the different physical assumptions between our models and those of \citet{Elstner+Gressel12}.
However, there is another important caveat, 
namely that the separation of the magnetic field into random and mean components 
may be sensitive to the method of averaging.
The mean-field dynamo theory on which our models are founded assumes ensemble averaging,
while \citet{Gressel+08b} use horizontal averaging.

\subsection{Scaling relation proportionality coefficients}\label{sec:coef}
Thus far we have focused on the scaling relation exponents.
We now make an order-of-magnitude estimate of the proportionality coefficients
in the magnetic field strength vs.~$\Sf$ relations.
This is important for checking that the magnitude of the magnetic field strengths
predicted by the model roughly agree with observational
and theoretical expectations.
We must first fix the values of the other observables;
this is done by choosing a typical value 
based on data for the galaxies M31, M33, M51 and NGC~6946,
which are the galaxies studied in Paper~II (in preparation).
We choose $\Omega=20\kmskpc$, $\Sigma=5\Msunpcpc$, and $\St=175\Msunpcpc$, respectively.
In addition we choose $T=10^4\K$ and $q=1$.
Further, 
the adjustable parameters of the model are chosen to be values
that generally produce good agreement with the velocity dispersion and magnetic field data in the preliminary analysis of Paper~II:
$\zeta=10$, $\beta=8$, $C_\alpha=4$, and $K=0.3$.

The scaling law can be written as
\begin{equation}
  B=\Btilde\left(\frac{\Sf}{\Sigmatilde}\right)^j,
\end{equation}
where $\Sigmatilde=10^{-3}\Msunyrkpckpc$ is a typical value of the SFR surface density
and $\Btilde$ is the coefficient we want to estimate.
To calculate $\Btilde$, we also need a proportionality coefficient
for the KS relation $\Sf=C\Sg^N$.
Using the data for $\Sf$ and $\Sg$ used in Paper~II, 
and assuming $N=1.4$, we obtain $C\approx10^{-15}$ in cgs units.
As an alternative, we estimated the coefficient from \citet{Leroy+08},
who found that the star formation efficiency is, on average, 
$\sim3\times10^{-10}\yr^{-1}$, for $\Sg\approx5$--$10\Msun\pc^{-2}$.
For $N=1.4$, this leads to $C\approx10^{-16}$ in cgs units.
Then, with these values, we obtain the values in Table~\ref{tab:coef}.
These rough estimates are consistent, at the order-of-magnitude level,
with those of \citet{Heesen+23}, 
who estimate a range $4.7$--$7.4\mkG$ for $\Btilde$.
Moreover, we generally find $b\iso>b\ani>\Bbar$, 
which is consistent with the usual ordering inferred by \citet{Beck+19}.
We note, however, that this ordering can be sensitive to the parameters of our model.

\begin{table}
\centering
\label{tab:coef}
\caption{Typical numerical coefficients $\Btilde_i$ for scaling relations of the form $B_i=\Btilde_i(\Sf/\Sigmatilde)^j$, where $\Sigmatilde=10^{-3}\Msunyrkpckpc$, 
for two different estimates of the numerical coefficient in the Kennicutt-Schmidt law relating $\Sf$ and $\Sg$ (see Section~\ref{sec:coef} for details).
Fitting data for the total magnetic field strength in multiple galaxies (estimated assuming energy equipartition with cosmic rays), 
\citet{Heesen+23} obtain $\Btilde$ in the range $4.7$--$7.4\mkG$.
Our model coefficients are generally of the same order of magnitude as theirs. All field strengths are in $\!\mkG$ and the KS relation coefficient is in cgs units.}
\begin{tabular}{@{}cccccccccc@{}}
\toprule
& \multicolumn{3}{c}{Model S} & \multicolumn{3}{c}{$\;$Model Alt1} &\multicolumn{3}{c}{$\;$Model Alt2} \\
KS coef.& $b\iso$ & $b\ani$ & $\Bbar$ & $\;\; b\iso$ & $b\ani$ & $\Bbar$ & $\;\; b\iso$ & $b\ani$ & $\Bbar$ \\ 
\midrule
$10^{-15}$ &10 &2  &1 &8  &3 &3 &12 &4  &3 \\ 
$10^{-16}$ &44 &16 &2 &20 &7 &6 &28 &10 &6 \\
\bottomrule
\end{tabular}
\end{table}

\section{Summary and conclusions}\label{sec:conclusions}
We have presented a model for the turbulence parameters
and magnetic field properties of disk galaxies that takes as input 
various observables.
The set of algebraic equations comprising the model 
can be solved semi-analytically 
using the source code linked in Section~\ref{sec:model}.
Solutions depend on the galactocentric radius 
and represent averages over the azimuthal and vertical coordinates.
The model rests on the assumption that the system is in a statistically steady state,
and any dependence on the cosmological redshift is neglected,
so our solutions are applicable to nearby galaxies
(though extending the model to include higher redshifts would be an interesting direction).
Figure~\ref{fig:flowchart_conceptual} summarizes the structure of the model.
A list of the quantities and parameters in the model 
and the equations for computing them can be found in Table~\ref{tab:params}.

\subsection{Scaling relations}
\label{sec:scaling_relations_summary}
We used the model to derive scaling relations for the gas scale height,
key turbulence parameters and magnetic field properties,
in terms of the observables.
These are relations of the form $X\propto x^ay^bz^c\cdots$, 
where $X$ is a quantity of interest,
$x$, $y$ and $z$ are observable quantities, and $a$, $b$ and $c$ are constants.
Scaling relations are useful for placing priors on missing information 
in datasets and for providing physical insight, and are ubiquitous in astrophysics.
The scaling relations can be derived analytically,
but we provide a link to a general numerical tool in Section~\ref{sec:model}.
To reduce the solutions to scaling relations, 
we focused on certain plausible asymptotic regimes. 
Most importantly, we assumed that turbulence is driven purely by isolated SNe
and that mean (large-scale) vertical and radial gas motions can be neglected.
In deriving scaling relations for the mean magnetic field strength, 
we assumed that the Coriolis number is small 
and that the dynamo is strong (Section~\ref{sec:scaling}), 
but these assumptions were not necessary 
for deriving scaling relations for the pitch angle of the mean field.

These scaling relations can be found in Tables~\ref{tab:scaling_int} and \ref{tab:scaling_B}.
Our fiducial model considers turbulence to be driven by isolated SNRs,
as they decelerate to the sound speed and merge with the ISM.
Given the lack of consensus about turbulence driving in the ISM, 
we also considered two simpler prescriptions for calculating the turbulence parameters, 
as summarized in Table~\ref{tab:models}.

The predictions in Tables~\ref{tab:scaling_int} and \ref{tab:scaling_B} 
do not consider possible correlations 
between the observed quantities on the right side of the scaling relations.
Given the well-known empirical relation between $\Sf$ and $\Sg$ (the Kennicutt-Schmidt law),
we used $\Sf\propto\Sg^{N}$ with $N=1.4$ to eliminate the dependence on $\Sg$, 
and the results are presented in Tables~\ref{tab:scaling_int_KS} and \ref{tab:scaling_B_KS}.

The theoretical scaling relations 
in Tables~\ref{tab:scaling_int_KS} and \ref{tab:scaling_B_KS}
were then compared with empirical scaling relations in the literature 
(Section~\ref{sec:comparison_observations}).
The level of agreement between the model predictions and observations
is remarkably good for our fiducial model,
Model~S,
given the various theoretical and observational uncertainties.
The level of agreement is generally poorer for the alternative turbulence prescriptions we tried (Models~Alt1 and Alt2).
This can be taken as further evidence that turbulence driving in nearby galaxies
is dominated by SN feedback. 
It also suggests that assuming that turbulence is driven 
at the disk scale height can lead to incorrect results,
and points to a need for modeling turbulence driving in more detail, 
taking into account the dynamics of SNRs.

\subsection{Limitations and future work}
\label{sec:limitations}
Our model should be thought of as an adaptable tool for understanding galactic magnetic fields
rather than a fixed set of formulae. 
This tool could be extended by including additional physical effects.
Outflows and SBs (which tend to cause outflows) are already included in the most general version 
of our model, summarized in Appendix~\ref{sec:gen},
but including these effects introduces extra parameters that are challenging to constrain.
SBs may be unlikely to dominate turbulence driving 
because they tend to blow out of the disk, 
which limits the amount of energy they transfer to the ISM \citep{Boomsma+08,Chamandy+Shukurov20}.
Likewise, rough estimates suggest that mean outflow speeds may often be too small
to strongly affect 
mean-field dynamo action \citep{Chamandy+16}.
However, such findings are preliminary and more work is needed on the roles of SBs and outflows
in turbulence driving and magnetic field evolution.
Furthermore, our model does not include radial inflow \citep[e.g.][]{Schmidt+16,Trapp+22}, 
which may affect the mean-field dynamo \citep{Moss+00} 
and play a role in driving interstellar turbulence, 
though primarily at high redshift \citep{Krumholz+18}. 

On the MHD side, 
we have not included the turbulent tangling of the mean field to produce random field,
nor additional effects~-- still not very well-understood~-- involving the influence of
the random field on the mean-field dynamo 
\citep[e.g.][]{Chamandy+Singh18,Gopalakrishnan+Subramanian23}.
Furthermore, the equations for the turbulence parameters, gas scale height, etc., 
do not take into account the magnetic field.
This is hard to avoid given the lack of current knowledge about such feedback effects,
and attempting to include them would make the model more complicated without providing much benefit,
given the various uncertainties.
Nevertheless, such effects may be important \citep[e.g.][]{Evirgen+19}.
In some cases, they can be roughly included in our existing model by choices of parameter values;
for example, the effect of the magnetic pressure on the gas scale height can be included heuristically
by increasing the value of $\zeta$ in equation~\eqref{h}.

Our model does not consider cosmological evolution,
but there is scope for extending the model to include high redshift galaxies
\citep{Schleicher+Beck13,Krumholz+18,Schober+23}.

There is 
scope for combining our model with galaxy models 
that rely on certain magnetic field parameters as input 
in order to calculate those parameters self-consistently.
For example, as mentioned in the introduction, 
the galaxy model of \citet{Wang+Lilly22} depends critically
on the exponent in the scaling relation between total magnetic field strength $B$
and SFR surface density $\Sf$,
as do theoretical models of the infrared-radio correlation \citep[e.g.][]{Schleicher+Beck13}.
In our model, there is~--- in the strict sense, at least~--- no scaling relation between 
total field strength $B$ and $\Sf$ because $B=(b\iso^2+b\ani^2+\Bbar^2)^{1/2}$, 
with each component having its own separate scaling relation 
(for $b\ani$ and $\Bbar$, that too only in certain plausible asymptotic limits).
However, the exponents in each of these relations lie in the range $0.1$--$0.4$ for our fiducial model, Model~S (Table~\ref{tab:scaling_B_KS}),
and this range is broadly consistent with the values inferred from observations (Section~\ref{sec:j_disk} and the left panel of Fig.~\ref{fig:B_Sf}).
Thus, our model can be seen as a possible refinement to models that 
assume that $B$ simply scales with $B\eq$ (defined in equation~\ref{Beq}) without other scalings.
Even if $B/B\eq=\const$ is assumed, one still needs a relation between the rms turbulent speed $u$ and $\Sf$.
Our fiducial model, Model~S, predicts a very weak dependence of $u$ on $\Sf$, 
with power-law exponent in the range $-0.03$ to $-0.02$ (Table~\ref{tab:scaling_int_KS}).
Here again, our model, which includes a detailed model of turbulence driving by SNe, 
can be seen as a possible refinement to other approaches.

Modeling scaling relations can be complicated by correlations between the observables.
While we used the KS relation to remove $\Sigma$ as an independent variable,
we made no attempt to incorporate the resolved star-forming main sequence relation between $\Ss$ and $\Sf$,
for instance.
If this could be used to eliminate $\Ss$, 
it would affect the $\Sf$-dependence of the scaling relations derived.
Nor did we attempt to relate the angular rotation speed $\Omega$ with the disk surface density
$\St$, for example, though they are not completely independent.
Another caveat is that parameters like $\zeta$ may vary somewhat between galaxies,
which would introduce scatter.

Observationally derived scaling relations for magnetic fields are now plentiful
and will improve as new instruments like the Square Kilometre Array become available.
This increases the urgency of studying such scaling relations theoretically, 
and in our view various complementary approaches can be utilized.
For instance, one could make use of a population synthesis model
that solves for the magnetic fields of galaxies
using 
the output of
a semi-analytic model of galaxy formation as input \citep{Rodrigues+15,Rodrigues+19}.
Second, one could run several local ISM simulations to explore the parameter space,
building on the work begun by \citet{Elstner+Gressel12}.
Third, one could explore scaling relations using cosmological zoom MHD simulations \citep[e.g.][]{Pakmor+24}.
Given the heterogeneous nature of galaxies, 
such a multi-pronged statistical approach may be very useful
for learning about galactic magnetic fields and the various processes that shape them.

\section*{Acknowledgements}
We are indebted to Anvar Shukurov for providing detailed comments on the manuscript
and for ongoing discussions about galactic magnetism and interstellar turbulence.
We are grateful to the referee for a constructive report 
that led to improvements to the presentation.
We also thank Rainer Beck for insightful comments on both an early and a recent version of the manuscript,
and Jennifer Schober and Luiz Felippe S. Rodrigues for discussions.

\section*{Data Availability}
There is no additional data to report.

\footnotesize{
\bibliographystyle{aasjournal}
\bibliography{refs}{}
}

\appendix

\section{Generalizations of the model}
\label{sec:gen}
Above, we neglected galactic outflows.
Applying an expression for the mean outflow speed $U\f$ derived in \citet{Vaneck+15} (see also \citealt{Shukurov+06,Rodrigues+15}),
\citet{Chamandy+16} found outflows to affect 
negligibly the mean magnetic field properties for the five galaxies 
for which the data to compute this quantity was available.
Even so, outflows may still have important effects, 
and in Section~\ref{sec:outflow}, we present mean-field dynamo equations for the case $U\f\ne0$.

If turbulence driving by SBs is important, the mean magnetic field then also depends on the fraction of SNe that contribute to SBs, $f\SB$, 
the number of SNe per SB, $N\SB$, and the energy efficiency of SBs, $\epsilon$, 
which in turn may depend on each other and on other parameters 
in ways that are difficult to constrain with current knowledge.
Thus, above, we neglected SN clustering, i.e.~we adopted $f\SB=0$.
\citet{Chamandy+Shukurov20} find that this assumption 
does not drastically alter the values of the turbulence parameters, 
typically making $\tau$ about $2$--$3$ times smaller and $l$ about $2$ times smaller 
compared to the fiducial case $f\SB=3/4$, 
whereas $u$ is similar in the two cases.
Nevertheless, SBs may sometimes be important, 
so in Section~\ref{sec:SBs} we formulate the equations for $f\SB\ne0$.
The mean vertical outflow speed $U\f$ is likely affected by SBs, 
so may itself depend on $f\SB$, $N\SB$ and $\epsilon$.

\subsection{Including mean vertical outflows}\label{sec:outflow}
This section generalises the magnetic field model of Section~\ref{sec:magnetic_field} 
to include a galactic outflow, with outflow speed $U\f$.

Using an expression from \citet{Hollins+17} we can write
\begin{equation}
  \label{sigma_z_gen}
  \sigma_z\simeq \sigma_r\left[1+\frac{K_U U\f\tau}{l}\left(1+\frac{1}{(1+q\Omega\tau)^2}\right)\right]^{1/2},
\end{equation}
where $U\f$ is defined to be equal to the mean vertical outflow speed at the disk surface $z=\pm h$,
with $h$ the scale height of the gaseous disk, and $K_U$ a constant of order unity.
As it represents an average for the entire disk, 
the quantity $U\f$ should be taken as the mass-weighted, area-averaged outflow speed,
and has been estimated to be in the range $0.2$--$2\kms$ for spiral galaxies \citep{Shukurov+06}.

Substituting equations~\eqref{sigma_phi} and \eqref{sigma_z_gen} into the expressions for $b$ and $b\iso$, 
and setting $K_U=1$, we obtain
\begin{equation}
\label{b_ani_gen}
  b\ani\equiv \sqrt{b^2-b\iso^2}= \frac{1}{\sqrt{3}}
    \left[2q\Omega\tau\left(1+\frac{q\Omega\tau}{2}\right)+\frac{U\f\tau}{l}\left(1+\frac{1}{(1+q\Omega\tau)^2}\right)\right]^{1/2}.
\end{equation}

The mean magnetic field strength in the saturated state is now given by
\begin{equation}
  \label{Bbar_gen}
  \Bbar\equiv |\bm{\Bbar}|= K B\eq\,\chi(p)\left(\frac{D}{D\crit} -1\right)^{1/2}\frac{l}{h}\left(R_U +\uppi^2 R_\kappa\right)^{1/2}.
\end{equation}
where we make use of the Reynolds-type dimensionless number 
\begin{equation} \label{R_U}
  R_U\equiv \frac{U\f h}{\eta}.
\end{equation}
The critical dynamo number is now given by
\begin{equation}
  \label{Dc_gen}
  D\crit= -\left(\frac{\uppi}{2}\right)^5\left(1 +\frac{1}{\uppi^2}R_U\right)^2.
\end{equation}
Additionally, we have
\begin{equation}
  \label{chi}
  \chi(p)=\left(2-\frac{3\cos^2p}{2\sqrt{2}}\right)^{-1/2},
\end{equation}
where
\begin{equation}
  \label{p_B_gen}
  \tan p_B= \frac{\Bbar_r}{\Bbar_\phi}= \frac{\uppi^2 +R_U}{4R_\Omega} = -\frac{\uppi^2\,\tau\,u^2+6\,h\, U\f}{12\,q\,\Omega\, h^2},
\end{equation}
and $p_B$ is the pitch angle of $\bm{\Bbar}$, 
defined such that $-\uppi/2<p_B\le\uppi/2$ with $p_B<0$ for trailing spirals.
Note that $\chi\approx1$, depending only weakly on $p_B$.
Given the uncertainties of the model, this dependence is inconsequential, and thus we set $\chi=1$.
The quantity $K$ in equation~\eqref{Bbar_gen} is an adjustable parameter of the model.

\subsection{Including superbubbles} \label{sec:SBs}
This section summarizes the SN-driven turbulence model of \citet{Chamandy+Shukurov20}.
For an overall SN rate per unit volume $\nu$,
the rate per unit volume of isolated SNe is given by
\begin{equation}
  \label{nu_SN}
  \nu\SN=(1-f\SB)\nu, 
\end{equation}
where $f\SB$ is the fraction of SNe in SBs, 
and that of SBs is given by
\begin{equation}
  \label{nu_SB}
  \nu\SB=\frac{f\SB\nu}{N\SB},
\end{equation}
where $N\SB$ is the mean number of SNe in a given SB.
\citet{Higdon+Lingenfelter05} estimate that the fraction of SNe occurring in OB associations is $\sim 3/4$ 
for the Milky Way,
which suggests $f\SB\sim3/4$ for our Galaxy.

\subsubsection{Turbulent correlation length $l$}\label{sec:l_gen}
The turbulent correlation length is estimated as
\begin{equation}
  \label{l_gen}
  l= \left(\frac{\Gamma-1}{\Gamma}\right)C_l l\SB\left(\frac{1+(l\SN/l\SB)\dot{\E}\inj\SN/\dot{\E}\inj\SB}{1+\dot{\E}\inj\SN/\dot{\E}\inj\SB}\right),
\end{equation}
where $l\SN$ and $l\SB$ are the driving scales for isolated SNe and SBs, respectively,
and $\dot{\E}\inj\SN$ and $\dot{\E}\inj\SB$ are their respective energy injection rates.
For Kolmogorov turbulence, $\Gamma=5/3$, 
and $C_l$ is a constant of order unity that comes from turbulence theory -- below we adopt $C_l=3/4$ \citep{Monin+Yaglom75}.

For SBs, we identify the driving scale of turbulence with the final radius reached by an SB in the midplane of the galaxy,
\begin{equation}
  \label{l_SB}
  l\SB\approx    \min\left[R\SB(t\sound\uSB), \lambda h\right],
\end{equation}
where the first case corresponds to deceleration to $c\sound$ and the second case to blowout from the disk.
Here $t\sound\uSB$ is the SB age for which the SB expansion has slowed to the ambient sound speed (if it has not blown out), 
$\lambda$ is a parameter of order unity, and
\begin{equation}
  \label{SB_Rs}
    R\SB(t\sound\uSB) = 0.53\kpc\, \epsilon_{0.1}^{1/3}N_{100}^{1/3}E_{51}^{1/3}n_{0.1}^{-1/3}c_{10}^{-2/3},
\end{equation}
where $0.1\epsilon_{0.1}$ 
is the fraction of the SB energy that is mechanical and $N_{100}=N\SB/100$.

The ratio of the rates of energy per unit volume injected by isolated SNe and SBs is
$l^3\SN\nu\SN/l^3\SB\nu\SB$, which gives
\begin{equation}
  \label{Eratio}
    \frac{\dot{\E}\inj\SN}{\dot{\E}\inj\SB}= 
      \begin{dcases}
                     0.63 \left(\tfrac{3(1-f\SB)}{f\SB}\right)\epsilon_{0.1}^{-1}
                     E_{51}^{-1/17}n_{0.1}^{-2/17}c_{10},
                      &\qquad \qquad  \mbox{if } t\sound\uSB\le t\blowout\uSB;\\
                     1.47 \left(\tfrac{3(1-f\SB)}{f\SB}\right)
                     N_{100}E_{51}^{16/17}n_{0.1}^{-19/17} c_{10}^{-1}\lambda^{-3}h_{0.4}^{-3},
                      &\qquad \qquad \mbox{if } t\blowout\uSB< t\sound\uSB.
      \end{dcases}
\end{equation}
Here $t\blowout\uSB$ is the age of the SB when it blows out of the disk (if it in fact blows out).
The similarity solution for SBs yields, from $\dot{R}\SB( t\sound\uSB)=c\sound$,
\begin{equation}
  \label{SB_ts}
  t\sound\uSB = 31\Myr\; \epsilon_{0.1}^{1/3}N_{100}^{1/3}E_{51}^{1/3}n_{0.1}^{-1/3}c_{10}^{-5/3}.
\end{equation} 
From $R\SB(t\blowout\uSB)=\lambda h$, we find 
\begin{equation}
  \label{tb}
  t\blowout\uSB= 15\Myr\; \epsilon_{0.1}^{-1/2}N_{100}^{-1/2}E_{51}^{-1/2}n_{0.1}^{1/2}\lambda^{5/2}h_{0.4}^{5/2},
\end{equation}
where $h_{0.4}= h/(0.4\kpc)$.

\subsubsection{Root-mean square turbulent velocity $u$}
\label{sec:u_gen}

Applying the method briefly outlined in Section~\ref{sec:u} to the case $f\SB\ne0$,
we obtain the general expression
\begin{equation}
  \label{u_gen}
  u= \left[ \frac{4\uppi}{3}l c\sound^2\nu\left(
           (1-f\SB)l\SN^3+\frac{f\SB}{N\SB}l\SB^3
           \right)
     \right]^{1/3}.
\end{equation}

\subsubsection{Turbulent correlation time $\tau$}
\label{sec:tau_gen}

The quantity $\tau\renov$ is the time for the flow to renovate due to the passage of an SN or SB blast wave.
The renovation rate is equal to the sum of the rates from isolated SNe and SBs, 
so the renovation time is given by
\begin{equation}
  \label{tau_renov_gen}
  \tau\renov= \left(\frac{1}{\tau\SN\ren} +\frac{1}{\tau\SB\ren}\right)^{-1}.
\end{equation}
A generalisation of equation~\eqref{tau_renov_noSBs} to $f\SB\ne0$ gives
\begin{equation}
  \label{tau_SN_renov}
    \tau\SN\ren= 6.8\Myr\; \left(\frac{1}{4(1-f\SB)}\right)
                      \nu_{50}^{-1}E_{51}^{-16/17}n_{0.1}^{19/17}c_{10},
\end{equation}
where $\nu_{50}=\nu/(50\kpc^{-3}\Myr^{-1})$.
The renovation time for SBs is equal to $3/(4\uppi l^3\SB\nu\SB)$, which gives
\begin{equation}
  \label{tau_SB_renov}
    \tau\SB\ren=  
      \begin{dcases}
        4.3 \Myr\;\left(\tfrac{f\SB}{3/4}\right)^{-1}\nu_{50}^{-1}
                 \epsilon_{0.1}^{-1}
                 E_{51}^{-1} 
                 n_{0.1}
                 c_{10}^2,
                 &\qquad \qquad \mbox{if } t\sound\uSB\le t\blowout\uSB;\\
        9.9 \Myr\;\left(\tfrac{f\SB}{3/4}\right)^{-1}\nu_{50}^{-1}
                 N_{100}
                 \lambda^{-3} 
                 h_{0.4}^{-3},
                 &\qquad \qquad \mbox{if } t\sound\uSB> t\blowout\uSB.
      \end{dcases}
\end{equation}

\end{document}